%% file: main.tex
\documentclass[aps,prl,reprint,superscriptaddress,longbibliography
]{revtex4-2}
\usepackage{amsmath,amssymb,bm}
\usepackage{graphicx}
\usepackage{lipsum} 
\usepackage{lipsum} 
\usepackage{float}
\usepackage{xcolor}
\usepackage{amsmath}
\usepackage{amsmath,amsthm,amssymb}

\usepackage{physics}
\usepackage[caption=false]{subfig}
\usepackage{graphicx}
\usepackage{dcolumn}
\usepackage{bm}
\usepackage{hyperref}

\makeatletter
\def\@seccntformat#1{\csname the#1\endcsname.\quad}
\makeatother
\begin{document}

\preprint{APS/123-QED}

\title{Scalable accuracy gains from postselection in quantum error correcting codes}

\author{Hongkun Chen}

\affiliation{%
Department of Electrical and Computer Engineering, Princeton University, Princeton, NJ 08544, USA
}
\author{Daohong Xu}

\affiliation{%
Department of Electrical and Computer Engineering, Princeton University, Princeton, NJ 08544, USA
}
 \author{Grace M. Sommers}
 \affiliation{Physics Department, Princeton University, Princeton, NJ 08544, USA}
 \author{David A. Huse}
 \affiliation{Physics Department, Princeton University, Princeton, NJ 08544, USA} 
 \author{Jeff D. Thompson}
 \affiliation{%
Department of Electrical and Computer Engineering, Princeton University, Princeton, NJ 08544, USA
}
\author{Sarang Gopalakrishnan}
\affiliation{%
Department of Electrical and Computer Engineering, Princeton University, Princeton, NJ 08544, USA
}

\date{\today}

\begin{abstract}

Decoding stabilizer codes such as the surface and toric codes involves evaluating free-energy differences in a disordered statistical mechanics model, in which the randomness comes from the observed pattern of error syndromes. We study the statistical distribution of logical failure rates across observed syndromes in the toric code, and show that, within the coding phase, logical failures are predominantly caused by exponentially unlikely syndromes. Therefore, postselecting on not seeing these exponentially unlikely syndrome patterns offers a scalable accuracy gain. In general, the logical error rate can be suppressed from $p_f$ to $p_f^b$, where $b \geq 2$ in general; in the specific case of the toric code with perfect syndrome measurements, we find numerically that $b = 3.1(1)$. Our arguments apply to general topological stabilizer codes, and can be extended to more general settings as long as the decoding failure probability obeys a large deviation principle. 
\end{abstract}

\maketitle

\section{Introduction}
Large-scale quantum computation requires fault-tolerance~\cite{gottesman2009introductionquantumerrorcorrection}. 
A key question for near-term quantum computation---where the number of good qubits is a limiting resource---is how to efficiently protect and manipulate quantum information in a fault-tolerant way on a fixed qubit budget. A strategy that is often used in practice is \emph{postselection}: because detecting errors is easier than correcting them, one can improve an algorithm's performance by repeating it and aborting trials in which errors are detected~\cite{google2021exponential, PhysRevX.11.041058, sivak2023real, bluvstein2024logical}. In near-term applications, postselection is a valuable error-mitigation technique~\cite{RevModPhys.95.045005, aharonov2025importance}. However, if the abort rate per measurement is finite, postselection is not scalable: only an exponentially small fraction of trials are kept. Several strategies have been developed to leverage postselection in scalable circuits, including syndrome extraction with verified ancillas \cite{SFCS.1996.548464, knill2004faulttolerantpostselectedquantumcomputation}, constructing cluster states from fixed-size resource states using fusion operations \cite{Bartolucci2023} and magic state distillation \cite{Bravyi_2005, Bravyi_2012}. In these approaches, the exponential overhead of postselection is mitigated by applying postselection to fixed-size chunks of the circuit.

In the present work, we instead consider directly postselecting on a circuit whose size scales with the code distance $d$. 
This approach was introduced in Ref.~\cite{smith2024mitigating}, which pointed out that postselection yields scalable accuracy gains for the surface code (see also Refs.~\cite{PRXQuantum.5.010302, english2024thresholdspostselectedquantumerror, Gidney2025}). 
The arguments of Ref.~\cite{smith2024mitigating} pertain to the surface code in the limit of physical error rate $p \to 0$. Here, we consider the more general setting of nonzero $p$, and present a statistical mechanics perspective that generalizes naturally beyond the surface code, and provides an intuitive argument for why postselection is useful. 

The decoding problem for a general stabilizer code maps to a disordered statistical mechanics problem, in which the randomness comes from the observed error syndrome, i.e., the pattern of violated error checks~\cite{Dennis_2002, Chubb_2021, PhysRevX.2.021004, Wang_2003, Katzgraber_2009, wu2025manybodyquantumchaostime}. In these mappings, the likelihood ratio between logical sectors is associated with a free-energy difference $\Delta F$, which in the toric code is between periodic and antiperiodic boundary conditions. We argue that, for error rates below the error correction threshold, at small $|\Delta F|$ the probability distribution $P(\Delta F)$ obeys a large deviation form~\cite{varadhan2010large}, $P(\Delta F = s d) \sim \exp(-I(s) d)$  
(e.g., $d$ is the linear size of the system in the examples of the toric or surface codes). Generic convexity properties of this distribution imply that the probability of incorrectly decoding a typical syndrome is exponentially lower (in $d$) than that of encountering a rare dangerous syndrome with $|\Delta F|$ near zero that is likely to be decoded incorrectly. 
This exponential separation implies that even weak postselection (where one discards only those exponentially few trials with small $|\Delta F|$) can decrease the logical failure rate by a factor that scales exponentially in $d$: 
thus one can enhance the effective code distance by a finite $O(1)$ factor. For nonzero but subthreshold error rates, we argue on convexity grounds that this factor is always at least $2$, but is generically larger: our numerical results on the toric code suggest a distance enhancement of a factor of $3.1(1)$. A striking consequence of our results is
that for certain tasks, strategies with postselection can scalably outperform those without. We check these arguments by numerically evaluating $P(\Delta F)$ for the toric code under independent $X$ and $Z$ errors and perfect measurements, using an optimal decoder, as well as for a more realistic circuit-level noise model under a minimum-weight perfect matching decoder~\cite{Gidney2025}.

\section{Model and setup}
We begin by reviewing the optimal decoding problem for the toric (or surface) code, and its mapping to a statistical mechanics model. For simplicity, we consider an error model of independent bit-flip ($X$) and phase-flip ($Z$) errors: since these are CSS codes, if the two types of errors occur independently, they can also be decoded independently, so without loss of generality we focus on bit-flip errors. In the standard representation of the toric code, qubits are placed on the links of a square lattice. Bit-flip errors are detected by checks (syndromes) that are products of $Z$ operators around plaquettes of the lattice. Errors form strings of contiguous flipped qubits, and the checks detect the endpoints of these strings.
Error correction consists of matching these endpoints (by guessing a pattern of error strings consistent with them, and applying bit-flips to join them along the guessed strings); it fails only if the true error string and the correction string combine to form a noncontractible loop around the torus. Thus, given a reference error string $E$ that is consistent with the observed syndrome, all other errors $E'$ consistent with the same syndrome can be classified into homology classes, depending on whether $E$ and $E'$ form a noncontractible loop around the torus. Error correction aims to use syndrome data to guess the homology class of $E$ correctly.

Assuming perfectly reliable measurements, the decoding problem for the toric code maps to a statistical mechanics model, the random-bond Ising model (RBIM)~\cite{Dennis_2002}, as follows. 
Checks map to Ising spins $s_i$, while physical qubits, labeled as $l_{ij}$, become Ising  couplings $J_{ij}$ between spins. For error chain $E$, the statistical mechanics Hamiltonian is
\begin{equation}
    H_E = -\sum_{\langle i, j\rangle}J_{ij}s_i s_j, ~~~~\text{where}~J_{ij}=\begin{cases}
    +1~~~~\text{if $l_{ij}\notin E$}\\
    -1~~~~\text{if $l_{ij}\in E$}
    \end{cases}.
\end{equation}
The Nishimori condition,
$P(J_{ij}=+1) =e^{2\beta}P(J_{ij}=-1)$, 
defines an inverse temperature $\beta$ set by the error probabilities, $P(J_{ij})$.
The total probability of errors within the same homology class as $E$ is proportional to the partition function ${\cal{Z}}(\beta,E)$. When the density of errors is below threshold, the statistical mechanics model is in its low-temperature ferromagnetic phase.

\begin{figure*}[t]
    \centering
    \includegraphics[width=\textwidth]{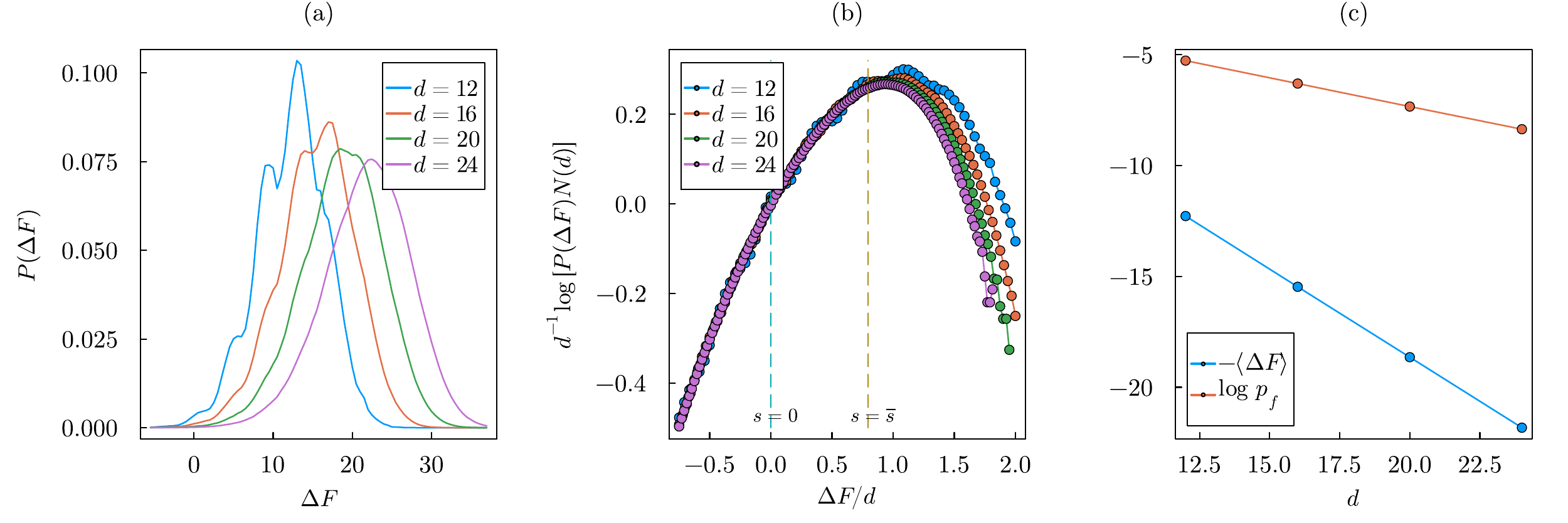}
    \caption{\label{fig: PDF scaling + ratio}
    Numerical results for $P(\Delta F)$ in the $\pm J$ RBIM (toric code) with $p = 0.06 < p_c \approx 0.109$.
    (a) $P(\Delta F)$ for different distances $d$. The peaks move to larger $\Delta F$ as $d$ increases, while error chains with $\Delta F$ near zero constitute only an exponentially small fraction of all error chains.
    (b) {Verification of large deviation scaling. After rescaling by $d$ and normalizing with $N(d)$, defined as the fitted value of $P(\Delta F=0)$, the curves for different system sizes collapse, consistent with the expected large deviation form. The scaling holds for $|s|<\overline{s}$, while for $|s|>\overline{s}$ the distribution exhibits a different asymptotic behavior}.  
    (c) Scaling of $\log p_f$ and $-\langle \Delta F\rangle$ with code distance $d$; $p_f$ is the logical failure probability. Both quantities decrease linearly with $d$, but with different slopes: $-\langle \Delta F\rangle$ decreases faster than $\log p_f$. We estimate that $\Delta F = F_0 + \overline{s} d$, where $\overline{s}=0.80(1)$, while $\log p_f = L_0 - I(0) d$, where $I(0)=0.26(1)$. Thus typical error configurations contribute negligibly to logical failure for large $d$, allowing efficient postselection.
}
\end{figure*}

Given the $J_{ij}$ arising from a given error chain $E$, the free-energy difference $\Delta F$ between the two homology classes relative to $E$ is defined as 
\begin{equation}
    \label{eq:Delta F def}
    \beta\Delta F=\log(\cal{Z}_{\text{PBC}}/\cal{Z}_{\text{APBC}})~,
\end{equation} 
where $\cal{Z}_{\text{PBC}}$ and $\cal{Z}_{\text{APBC}}$ are the partition functions under periodic and antiperiodic boundary conditions, respectively. (These two homology classes differ by a relative domain wall; domain wall in a random two-dimensional magnet can be mapped onto that of a directed polymer in a random environment~\cite{PhysRevLett.54.2708}.) Since $\Delta F$ depends on the error configuration, it is a random variable which takes the same value on all error configurations that produce the same syndrome and are in the same homology class.  This immediately implies the Nishimori symmetry constraint on the distribution $P(\Delta F)$:
\begin{equation}
    \label{eq:P(Delta F) constraint}
    P(\Delta F)=P(-\Delta F)e^{\beta \Delta F}~.
\end{equation}
Since every RBIM on the Nishimori line is associated with optimal decoding of a toric code with some error model \cite{suppmat}, Eq. \eqref{eq:P(Delta F) constraint} holds regardless of whether $P(J_{ij})$ is continuous or discrete. 
We rescale the $J_{ij}$ such that $\beta = 1$, and therefore use $\beta \Delta F$ and $\Delta F$ interchangeably in what follows. $Z$ is computed using the {time-evolving block decimation (TEBD)} method implemented with the ITensors.jl package~\cite{itensor, itensor-r0.3}; for details see~\cite{suppmat}.

We determine the logical failure probability as follows. We first sample an error chain $E$ from the distribution over errors. This is the ``ground-truth'' error and generates a particular syndrome (which is experimentally accessible). The optimal decoder computes the free energies with errors in both homological classes consistent with the observed syndrome, and infers that the error chain belonged to the class with the lower free energy. (Decoders lack access to the true error chain, and therefore do not know the sign of $\Delta F$; however, they can estimate their confidence in the error correction operation based on $|\Delta F|$.) Decoding fails when the {inferred homology class does not match that of $E$}, 
i.e., when $\Delta F < 0$. Therefore, the logical failure probability is

\begin{equation}
    \label{eq: logical failure rate}
    p_{f}=\int_{-\infty}^{0}P(\Delta F)\dd \Delta F=\int_{-\infty}^\infty\frac{P(\Delta F)}{1+e^{|\Delta F|}}\dd \Delta F~.
\end{equation}

\par 
\section{Large deviation analysis}
We now discuss the asymptotics of $P(\Delta F)$. Since this is the free-energy cost of a domain wall in a random ferromagnet, it can be parameterized as $\Delta F = s d$, where $s$ is an intensive random variable, the line tension of the domain wall. This free-energy cost obeys a large deviation principle (LDP):  For $|s|$ less than its typical value, $P(\Delta F = s d) \sim \exp(-I(s) d)$, where $I(s)\equiv\lim_{d\rightarrow\infty}\frac{\log P(sd)}{d}$ is called a ``rate function''~\cite{varadhan2010large, carmona2004fluctuation, carmona2010directed, janjigian2015large}. Physically, the LDP arises because, for a domain wall to have an atypically low value of $\Delta F/d$, it must pass through a string of regions with atypical disorder configurations. Thus, $\Delta F$ is a sum of weakly correlated random variables, which generically obey a LDP.

Numerical results from the RBIM (toric code) are shown in Fig.~\ref{fig: PDF scaling + ratio}. The scaling collapse in Fig.~\ref{fig: PDF scaling + ratio}(b) illustrates that the LDP holds for small $|\Delta F|$. At the peak of the distribution, $I(\overline{s})=0$. The constraint Eq.~\eqref{eq:P(Delta F) constraint} implies a symmetry relation for $I(s)$:
\begin{equation}\label{eq: Is-symmetry}
    I(s)+s=I(-s)~.
\end{equation}
We assume two standard properties of the rate function $I(s)$, which are also apparent in the numerics: is monotonically decreasing for $s\le\overline{s}$ and convex.

{By the Laplace principle, the integral is controlled by the minimum of the rate function in a large deviation regime,}
\begin{equation}
    p_f(d)\sim \int_{-\infty}^{0} e^{-I(s)d}\,ds\sim e^{-I(0)d}~,
\end{equation}
showing that logical failures are dominated by atypical syndromes in the vicinity of $\Delta F=0$. Typical syndromes, corresponding to $|\Delta F|=\overline{s}d$, give an exponentially smaller contribution  $\sim e^{-I(-\overline{s})d}=e^{-\overline{s}d}$ to $p_f$. 

The basic idea behind scalable postselection is that by suppressing syndromes with $|\Delta F| \leq s d$ for $0<s < \overline{s}$, we can suppress the logical failure probability from $\sim e^{-I(0)d}$ to $\sim e^{-I(-s)d}$---which is exponentially smaller---while only aborting an exponentially small fraction of trials.
Thus, postselecting on not seeing atypically dangerous syndromes gives a maximal gain in the effective code distance that is upper-bounded by a {distance-independent} factor of $b \equiv I(-\overline{s})/I(0)$. It is straightforward to show that the convexity of $I(s)$ guarantees that $b \geq 2$ for any code where the failure probabilities follow an LDP, regardless of whether Eq.~\eqref{eq: Is-symmetry} holds. 

It is computationally hard to sample rare events near $\Delta F=-\langle\Delta F\rangle$; however, the symmetry of $I(s)$ lets us use $\overline{s}$ to evaluate $I(\overline{s})$. For the RBIM, we numerically estimate $\overline{s}$ from the slope of $\langle \Delta F\rangle$ versus $d$, and $I(0)$ from the slope of $\log p_f(d)$ versus $d$ [Fig.~\ref{fig: PDF scaling + ratio}(c)], and find $b=3.1(1)$.
This ratio $b \approx 3$ appears rather robust for the toric code, even if one changes the error model (e.g., by introducing some density of heralded errors~\cite{suppmat}). Interestingly, $b=3$ is the ratio that one would get by extrapolating the Tracy-Widom distribution, which governs \emph{typical} fluctuations of the domain wall free energy~\cite{tracy2009distributions} ($s - \overline{s} \sim d^{-2/3}$), and extrapolating it to the large deviation regime.

 \par
\section{Code-splitting}
Next, we discuss a surprising implication of the convexity relation $b \geq 2$: for certain tasks, postselection offers advantages even for a fixed spacetime budget. To illustrate this we consider protecting a known, and thus clonable, logical state $\ket{\psi}$ against noise. Using $N$ qubits, we can do this by encoding $\ket{\psi}$ in a $\sqrt{N} \times \sqrt{N}$ toric code, or by preparing $r$ copies of $\ket{\psi}$ and encoding each in a toric code with dimensions $\sqrt{N/r} \times \sqrt{N/r}$. When $r > 1$ {syndromes are extracted in parallel,} we estimate $|\Delta F|$ for each copy, keep the copy with the largest $|\Delta F|$, and discard the rest. Since $r = 1$ has the largest code distance, one might expect that splitting always increases the logical failure probability. However, this is not the case. {We first consider the case $r=2$. The joint distribution obeys
$P(\Delta F_1=s_1 d,\Delta F_2=s_2 d)\sim \exp\{-[I(s_1)+I(s_2)]d\}$. 
Decoding selects $\max(|s_1|,|s_2|)$, and a logical error occurs when the selected value is negative, i.e., $s_1<0$ with $|s_1|>|s_2|$ (or symmetrically). 
By the Laplace principle, the probability is dominated by the boundary $s_1=-s_2-\epsilon<0$, yielding
$p_{f}\sim \exp\{-[I(s_1)+I(-s_1)]d\}$, which is maximized at $s_1=0$ for convex $I(s)$, so}

\begin{equation}\label{split2}
    p_f^{(2)}\left(d\right) \sim\exp\left\{-\min_{s\ge0}[I(s)+I(-s)]\, d\right\}=e^{-2I(0)d}.
\end{equation}
Indeed, Eq.~\eqref{split2} also implies that asymptotically, two distance-$d$ codes with the postselection strategy described above have the same asymptotic distance as a single distance-$2d$ code.  For a code, such as the toric code, that has the code distance $d$ growing slower than linearly in the number of qubits, this provides the same logical failure rate while using fewer qubits. We emphasize that this result is not specific to the toric code but is a generic consequence of the LDP and convexity.

\begin{figure*}[t]
    \centering
    \includegraphics[width=\textwidth]{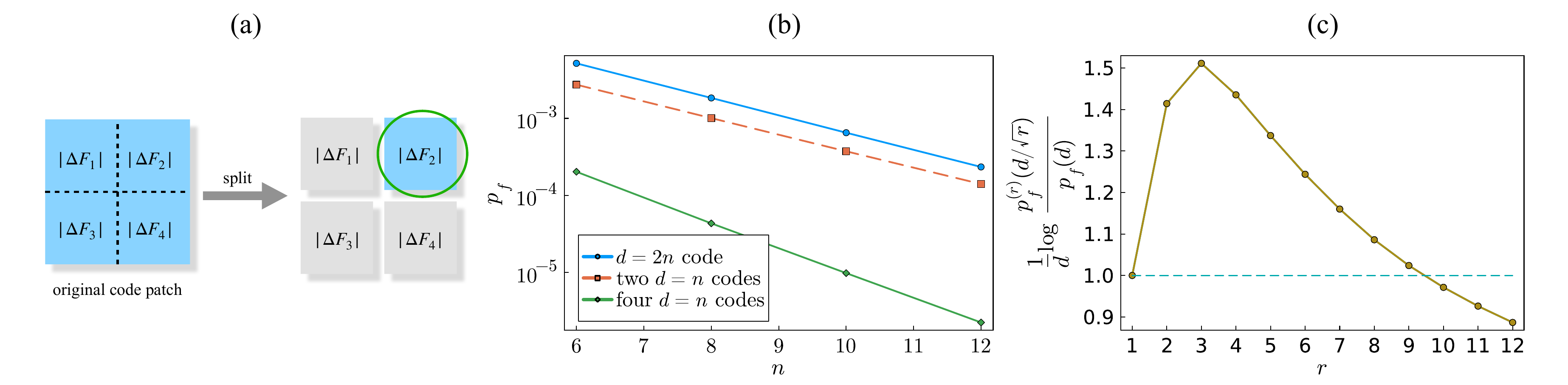}
    \caption{
(a) Illustration of code splitting for $r=4$: a code patch with $d = L$ is divided into four $d = L/2$ subpatches. During decoding, the patch with the largest $|\Delta F_i|$ (e.g., $|\Delta F_2|$) is selected.
\; (b) Comparison of logical failure probabilities for a single distance-$2n$ code, two distance-$n$ subcodes with postselection, and four distance-$n$ subcodes with postselection, for the $\pm J$ RBIM with disorder parameter $p=0.06$. The two $d=n$ codes exhibit the same scaling of logical failure probability as the single $d=2n$ code, while four $d=n$ codes decay exponentially faster with $n$. These results match our theoretical predictions.\; (c) Accuracy gain from code splitting as a function of the number of subcodes $r$. Numerics (panel~(b)) shows that the $r=4$ strategy enhances the effective distance by a factor $\approx 1.45$, consistent with rate-function predictions ($\approx 1.44$). 
\label{fig:code_splitting_combined}}
\end{figure*}

{More generally, the probability distribution of maximal $\Delta F_i$ among $r$ copies can be calculated via order statistics

~\cite{suppmat}}, as 
%
$p_f^{(r)}(d)\simeq\exp\{-\min_{s\ge0}[rI(s)+s]\, d\}$. 
The window of $r$ for which splitting yields a scalable advantage is plotted in Fig. \ref{fig:code_splitting_combined}. For the toric code, $r = 3$ seems optimal~\footnote{Splitting a toric code in three pieces requires one to be able to move and reassemble qubits, which is feasible in architectures like neutral atoms.}. Since the failure probability (for typical syndromes) of each subcode scales as $\sim\exp(-\overline{s} d )$ while the failure rate for the unsplit code (using the same number of qubits) is $\sim\exp(-I(0)d\sqrt{r})$, code-splitting cannot help when $r \geq b^2$, consistent with the numerics.

\par
\section{Postselection}
{The splitting protocol discussed above illustrates that postselection can yield gains even for a fixed space and time budget, by redistributing a fixed number of physical qubits across several distance-$d$ patches. A more realistic resource model allows for repeated trials, allowing one to trade time overhead for improved accuracy by aborting a trial if one finds a syndrome with an unacceptably small value of $|\Delta F|$.}
For scalability, we want postselection to be exponentially rare in $d$, so we consider aborting if we see syndromes with $|\Delta F| \leq s^* d$, where $s^* < \overline{s}$. For a single round of postselection, the abort rate is $\exp(-I(s^*) d)$, and the logical error rate conditional on not aborting is $\exp(-I(-s^*) d)$. 

For a given $d$ and $\mathcal{N}$ logical operations, the conditions for postselection to help can be estimated as follows. For concreteness, the estimate below will focus on the toric code with perfect syndrome measurements, but similar estimates can be made in other settings. Suppose we require that the probability of not aborting remains $O(1)$ as $d \to \infty$, i.e., that $\mathcal{N} \exp(-I(s^*) d) \alt 1$, assuming that one round of error correction happens between any two operations. 
This fixes $I(s^*) \simeq d^{-1} \log \mathcal{N}$, subject to the constraint that $s^* \geq 0$. This equation has nontrivial solutions when $d^{-1} \log \mathcal{N}$ is sufficiently small; the postselected logical failure rate in these cases scales as $\exp(-I(-s^*) d)$. In the limit of few logical operations ($\mathcal{N} = O(1)$), $s^* = \overline{s}$, since typical syndromes occur with high probability. Such solutions exist for any circuit of depth polynomial in $d$, since $\log \mathcal{N}/d \sim \log d /d$. However, for circuits of depth scaling exponentially in $d$, the amount of postselection that can be tolerated while maintaining an $O(1)$ abort rate decreases, until for deep enough circuits $s^* = 0$ and the gain from postselection is entirely lost. In addition, when the system is not in the decodable phase, $\overline{s}=0$. Thus, postselection with finite acceptance probability does not alter the error threshold, but effectively increases the code distance from $d$ to $I(-s^*)d/I(0)$, as conjectured in Ref.~\cite{english2024thresholdspostselectedquantumerror}. {The large-deviation analysis makes this explicit: above threshold one has $\overline{s}=0$, so any exponential accuracy gain would require postselecting on events in the far right tail of the distribution, of probability $\exp[-O(d^2)]$, and thus finite-acceptance postselection cannot raise the threshold.}

The strategy outlined above yields no advantage in the normal setting of fault-tolerant computation. In this setting, the relevant figure of merit is the minimum code distance required to complete an algorithm with $\mathcal{N}$ logical operations. Without postselection, reliable computation with distance-$d$ codes requires $\mathcal{N}e^{-I(0)d}\lesssim 1$. 
With postselection on $|\Delta F|\ge s^* d$, the logical error rate is further suppressed, seemingly reducing the distance overhead. Yet scalability requires a finite acceptance probability, which imposes $
    \mathcal{N}p_\text{abort}\simeq \mathcal{N}e^{-I(s^*)d}\lesssim 1$. 
Since $I(s^*)<I(0)$ for all $s^*>0$, the abort probability becomes the limiting factor: requiring an $O(1)$ abort rate imposes a more stringent constraint on $d$ than requiring an $O(1)$ logical failure probability.  Thus, postselecting on an entire circuit cannot be used to reduce overhead, but rather provides a scalable gain in \emph{accuracy}, in settings where the key objective is to minimize the probability of a logical failure. {While postselection on an entire circuit does not reduce overhead, recent works have shown that scalable postselection applied to logical Bell-pair preparation \cite{sunami2026entanglementboostinglowvolumelogical} and teleportation-based resource-generation subroutines \cite{staples2026scalablepostselectionquantumresources, birchall2026macromuxscalablepostselectionhighthreshold} can substantially reduce qubit and spacetime overheads.}

\par 

\section{Discussion}
\begin{figure}[t]
    \centering
    \includegraphics[width=0.85\linewidth]{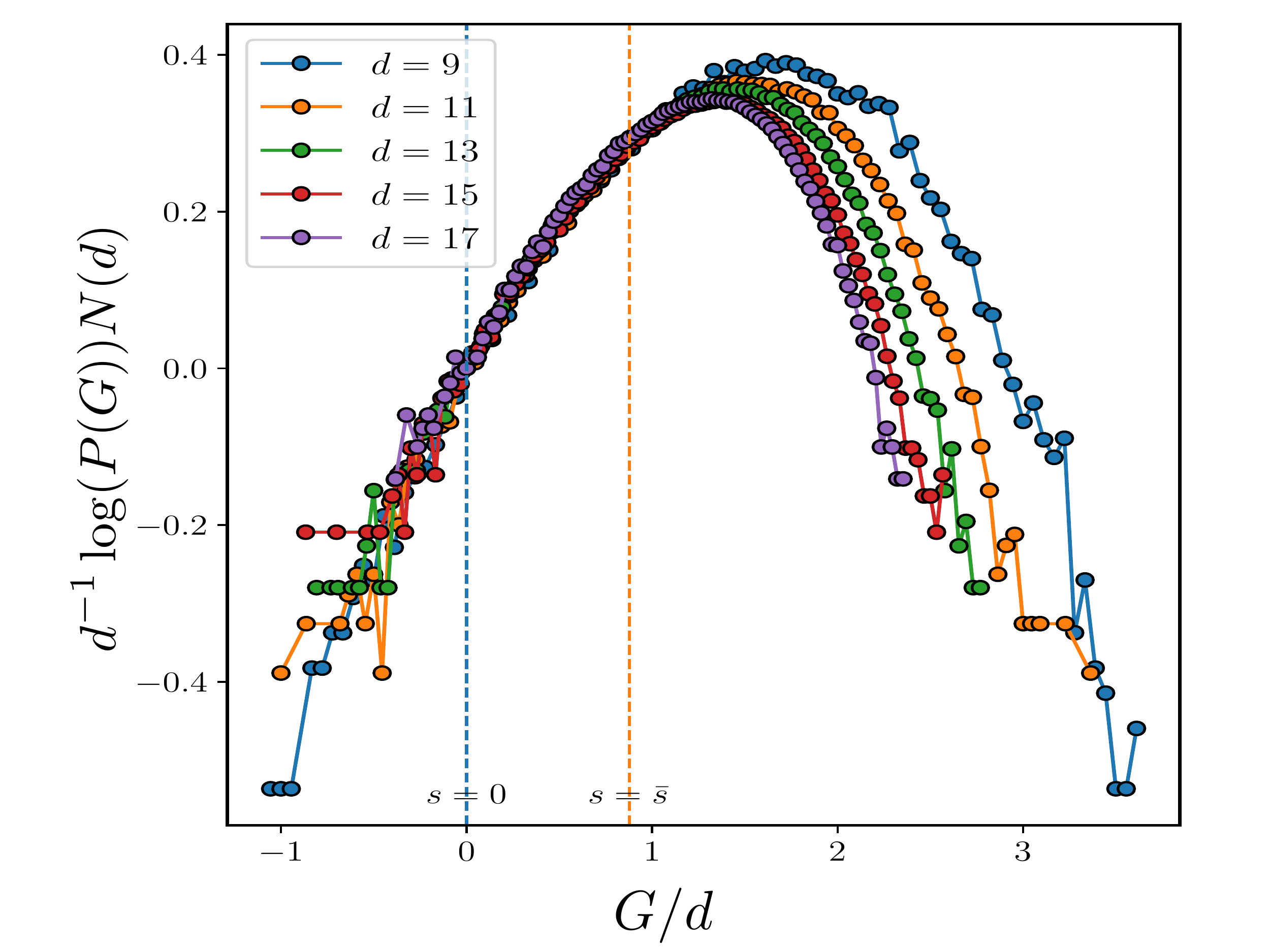}
    \caption{Verification of large deviation scaling for minimum-weight perfect matching (MWPM) surface code decoding under circuit-level noise, including depolarizing noise, measurement errors, and CNOT gate faults at rate $p=0.005$. The rescaled distribution of the complementary gap $G$, defined as the difference between the minimum weights in the two homology classes~\cite{suppmat}, exhibits clear data collapse across code distances, consistent with large deviation scaling within the range $|s|<\overline{s}$. 
    The behavior is analogous to that of the two-dimensional RBIM, demonstrating that the large deviation form persists under realistic circuit-level noise.}
    \label{fig: LD scaling under circuit-level noise}
\end{figure}

Above, we showed that postselection with parametrically low abort rates can almost triple the effective distance of the surface code. This gain in the effective distance is {asymptotic}, and follows from a large deviation principle (LDP) that we expect to hold much more generally. We focused above on the case with perfect syndrome measurements, but the extension to imperfect measurements is direct: this case maps to a different statistical mechanics model in spacetime~\cite{Dennis_2002}, and the logical operator is a directed polymer in a three-dimensional disordered medium. The free energy distribution of this polymer again takes a large deviation form. For this case, the optimal decoder is numerically intractable; however, the LDP also holds for suboptimal decoders, and we verify it numerically for the minimum-weight perfect matching decoder in Fig.~\ref{fig: LD scaling under circuit-level noise} (see also~\cite{suppmat}). 

More generally, in topological codes where the logical operator lives on a lower-dimensional surface of the system, the LDP is expected to apply for the same reason as in the surface code. 
We also present numerical evidence~\cite{suppmat} for this scaling in concatenated codes~\cite{PhysRevResearch.7.023086, Yadavalli_2025}. An interesting question for future work is whether an analogous LDP continues to hold in families of codes such as generic hypergraph product codes~\cite{tillich2013quantum} and good quantum LDPC codes~\cite{panteleev2022asymptotically}. 
{In principle, marginal decoding provides a well-defined framework to analyze decoding for arbitrary qLDPC codes~\cite{suppmat}.}
{From a mathematical perspective, the G\"artner--Ellis theorem \cite{gartner1977,ellis1984} implies that $\Delta F$ obeys a large-deviation principle with a convex rate function, provided that the scaled cumulant generating function (SCGF) $\lambda(k)$ is continuously differentiable~\cite{Touchette2009} Although we are currently unable to rigorously establish the existence and $C^1$ regularity of $\lambda(k)$ for general code families, the commonly expected exponential suppression of logical errors below threshold \cite{Fowler_2012} guarantees the convergence of the SCGF in part of the $k$ range and provide bounds on the corresponding SCGF sequence \cite{suppmat}. Therefore, under sufficiently regular behavior of the SCGF, the G\"artner--Ellis theorem provides a natural route to the LDP and convexity of the rate function.}

When the LDP does hold, an immediate consequence is that logical failures are dominated by rare events with $\Delta F\approx 0$. As these events get exponentially rare in the code distance, and lead to syndromes that can be identified as dangerous, there is scope for scalable postselection strategies that abort if they see dangerous syndromes. {However, computing the associated marginal probabilities is not expected to be efficient in general, and the practical applicability of this framework to general qLDPC codes therefore remains an open challenge. }

{During the submission of this manuscript, Ref.~\cite{lee2025efficientpostselectiongeneralquantum} appeared, which develops a practical method for implementing postselection in general qLDPC codes. Our work and Ref.~\cite{lee2025efficientpostselectiongeneralquantum} address complementary aspects of postselection: we focus on the large-deviation mechanism and use marginal decoding as an asymptotically optimal \emph{theoretical} decoder~\cite{PhysRevA.105.052446}, while Ref.~\cite{lee2025efficientpostselectiongeneralquantum} focuses on practical confidence estimation in realistic decoding settings and therefore emphasizes algorithmic and numerical implementations.}

\begin{acknowledgments}
We thank Will Staples for helpful discussions. H.C. and S.G. acknowledge support from NSF QuSEC-TAQS OSI 2326767. 
This research was supported in part by NSF QLCI grant OMA-2120757, including an Institute for Robust Quantum Simulation (RQS) seed grant.
\end{acknowledgments}


\bibliography{reference}

\pagebreak

\widetext

\newpage

\input{SM}
\end{document}

%% file: SM.tex
\makeatletter
\begin{center}
\textbf{\large Supplemental Materials: Scalable accuracy gains from postselection in quantum error correcting codes}

\vspace{3mm}

Hongkun Chen,\textsuperscript{1} Daohong Xu,\textsuperscript{1} Grace M. Sommers,\textsuperscript{2} David A. Huse,\textsuperscript{2} Jeff D. Thompson,\textsuperscript{1} and Sarang Gopalakrishnan\textsuperscript{1}

\vspace{2mm}

\textsuperscript{1}\textit{\small Department of Electrical and Computer Engineering,\\ Princeton University, Princeton, NJ 08544, USA}

\textsuperscript{2}\textit{\small Department of Physics, Princeton University, Princeton, New Jersey 08544, USA}

\makeatother


\makeatother

\end{center}

\setcounter{equation}{0}
\setcounter{figure}{0}
\setcounter{table}{0}
\setcounter{page}{1}
\makeatletter
\renewcommand{\theequation}{S\arabic{equation}}
\renewcommand{\thefigure}{S\arabic{figure}}

\section*{TEBD method for computing partition function of RBIM}
\textbf{Tensor network (TN) construction to represent $\mathcal{Z}$.} Computing the partition function of the two-dimensional random-bond Ising model (RBIM) is a fundamental yet notoriously challenging problem, particularly in the presence of disorder. For large system sizes, exact enumeration of all spin configurations becomes computationally intractable due to the exponential growth of the Hilbert space. To overcome this barrier, we employ the time-evolving block decimation (TEBD) algorithm, which efficiently approximates the thermal partition function as a tensor-network contraction. In this section, we detail the construction of TEBD for evaluating the partition function of the RBIM with arbitrary Ising couplings \( J_{ij} \) and specified boundary conditions.\par

To begin with, we consider a $d\times d$ square lattice with nearest Ising couplings and the periodic boundary condition (antiperiodic boundary can be realized by flipping the signs of couplings), and label all spins by their coordinates $(x, t)$, where $0\le x, t\le d-1$. The Hamiltonian $\mathcal{H}[\sigma]$ could be rewritten as
\begin{equation}
    \mathcal{H}[\sigma] = -\sum_{t=0}^{d-1}\sum_{x=0}^{d-1} J_{(x, t), (x+1, t)}\sigma_{(x, t)}\sigma_{(x+1, t)} - \sum_{t=0}^{d-1}\sum_{x=0}^{d-1} J_{(x, t), (x, t+1)}\sigma_{(x, t)}\sigma_{(x, t+1)}.
\end{equation}
The partition function $\mathcal{Z}$ can be expressed as 
\begin{equation}
    \mathcal{Z} = \sum_{\text{all}~\sigma_{(x, t)}} \prod_{x, t=0}^{d-1}e^{\beta J_{(x, t), (x+1, t)}\sigma_{(x, t)}\sigma_{(x+1, t)}}e^{\beta J_{(x, t), (x, t+1)}\sigma_{(x, t)}\sigma_{(x, t+1)}}.
\end{equation}

\begin{figure}[H]
    \centering
    \includegraphics[width=0.6\linewidth]{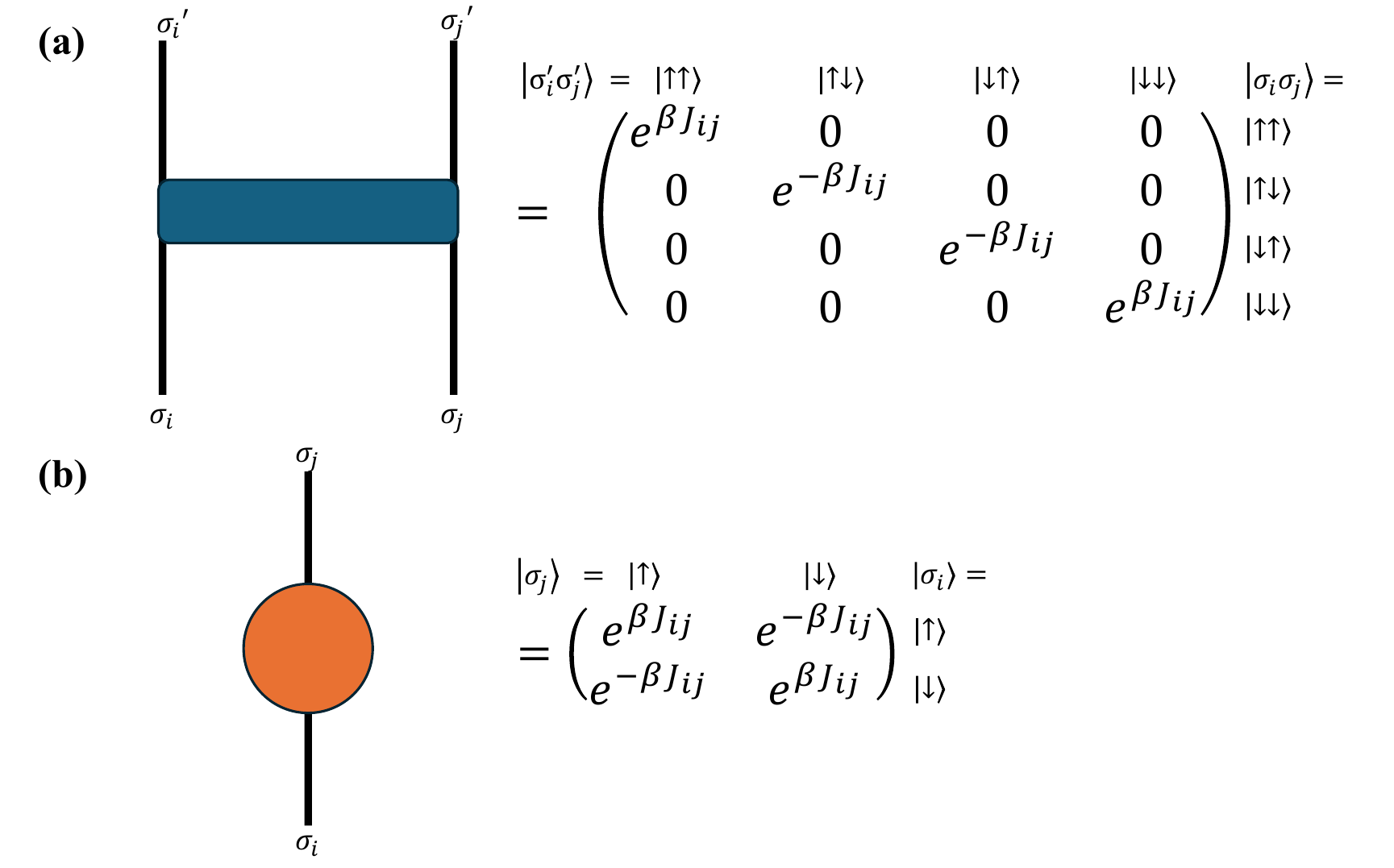}
    \caption{Graphic representations of matrices $\mathcal{B}_{\sigma_{i}'\sigma_{j}'}^{\sigma_i \sigma_j}$ and $\mathcal{S}_{\sigma_i}^{\sigma_j}$. They play as basic ingredients in constructing tensor network representation of partition function $\mathcal{Z}$. \textbf{(a)} Tensor $\mathcal{B}_{\sigma_{i}'\sigma_{j}'}^{\sigma_i \sigma_j}$ as a 4-leg tensor acting like a two-qubit gate. \textbf{(b)} Tensor $\mathcal{S}_{\sigma_i}^{\sigma_j}$ as a 2-leg tensor acting on a single qubit. }
    \label{fig:tensor1}
\end{figure}
By summing over $\sigma_{(x, t)}$ for certain $t$ and all $0\le x\le d-1$, one gets a tensor $\mathcal{T}_t$ with indices $\{\sigma_{(x, t)}\}$ and $\{\sigma_{(x, t+1)}\}$ as
\begin{equation}
    \mathcal{T}_{t}=\sum_{\text{all}~\sigma_{(x, t)}} \prod_{x=0}^{d-1}e^{\beta J_{(x, t), (x+1, t)}\sigma_{(x, t)}\sigma_{(x+1, t)}}e^{\beta J_{(x, t), (x, t+1)}\sigma_{(x, t)}\sigma_{(x, t+1)}}\\.
\end{equation}
$\mathcal{T}_t$ could be viewed as a transformation from indices $\{\sigma_{(x, t)}\}$ to $\{\sigma_{(x, t+1)}\}$ as a matrix, hence the partition function $\mathcal{Z}$ can be expressed as
\begin{equation}
    \mathcal{Z}=\Tr[\prod_{t=0}^{d-1}\mathcal{T}_t].
\end{equation}

\begin{figure}[t]
    \centering
    \includegraphics[width=1\linewidth]{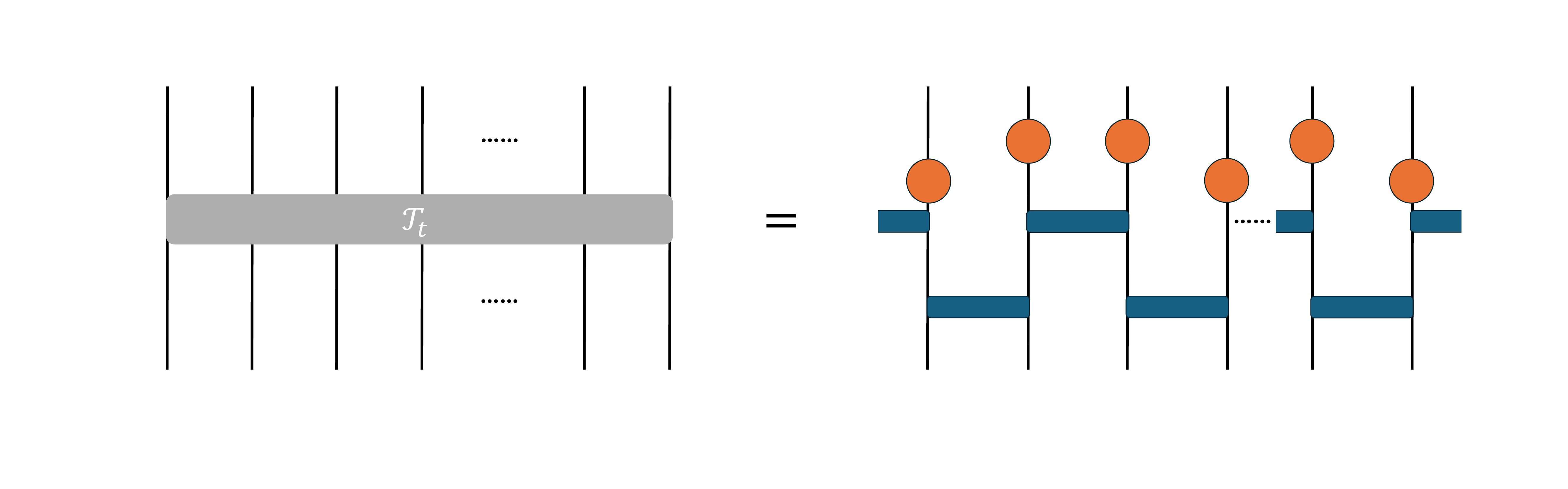}
    \caption{TN representation of $\mathcal{T}_t$ using local tensors $\mathcal{B}_{\sigma_{i}'\sigma_{j}'}^{\sigma_i \sigma_j}$ and $\mathcal{S}_{\sigma_i}^{\sigma_j}$. The overall action of $\mathcal{T}_t$ works like Trotterized time-evolution driven by two-qubit gates and single-qubit gates.}
    \label{fig:T_t}
\end{figure}

To find the tensor network representation of $\mathcal{T}_t$, we define the local tensors $\mathcal{B}_{\sigma_{i}'\sigma_{j}'}^{\sigma_i \sigma_j}$ and $\mathcal{S}_{\sigma_i}^{\sigma_j}$ as shown in Fig. \ref{fig:tensor1}(a) and Fig. \ref{fig:tensor1}(b), respectively. 
\begin{equation}
    \mathcal{B}_{\sigma_{i}'\sigma_{j}'}^{\sigma_i \sigma_j}=e^{\beta J_{ij}\sigma_i \sigma_j}\delta_{\sigma_i \sigma_{i}'}\delta_{\sigma_j \sigma_{j}'}, \quad \mathcal{S}_{\sigma_i}^{\sigma_j}=e^{\beta J_{ij}\sigma_i \sigma_j}.
\end{equation}
Then, as Fig. \ref{fig:T_t} illustrates, $\mathcal{T}_t$ is represented as (using Einstein summation convention)
\begin{equation}
    \mathcal{T}_t= \prod_{x=2m}\mathcal{B}_{\sigma_{(x, t)}\sigma_{(x+1, t)}}^{\sigma_{(x, t)}'\sigma_{(x+1, t)}'} \prod_{x=2m+1}\mathcal{B}_{\sigma_{(x, t)}'\sigma_{(x+1, t)}'}^{\sigma_{(x, t)}''\sigma_{(x+1, t)}''}\prod_{x}\mathcal{S}_{\sigma_{(x, t)}''}^{\sigma_{(x, t+1)}}.
\end{equation}

Since $\mathcal{B}_{\sigma_{i}'\sigma_{j}'}^{\sigma_i \sigma_j}$ acts like a two-qubit gate and $\mathcal{S}_{\sigma_i}^{\sigma_j}$ acts like a single-qubit gate, contracting such tensor network becomes a TEBD simulation of time evolution driven by single-qubit and two-qubit gates.

~\\ \par
\textbf{The accuracy of TEBD simulation. }To assess the accuracy of our TEBD implementation, we
benchmark it against a problem with a closed-form solution: the 
two-dimensional Ising model.  Onsager's result gives the exact
free-energy density in the thermodynamic limit as

\begin{equation}
f_{\infty}(K)
  = -\frac{\log 2}{2}
    \;-\;
    \log\!\bigl(\cosh 2K\bigr)
    \;-\;
    \frac{1}{2\pi}
    \int_{0}^{\pi}
       \log\!\Bigl(
             1+\sqrt{\,1-\kappa^{2}\cos^{2}\theta}
           \Bigr)\,d\theta,
\qquad
\kappa \;=\;\frac{2\sinh 2K}{\cosh^{2} 2K}.
\end{equation}

\begin{figure}[t]
    \centering
    \includegraphics[width=0.6\linewidth]{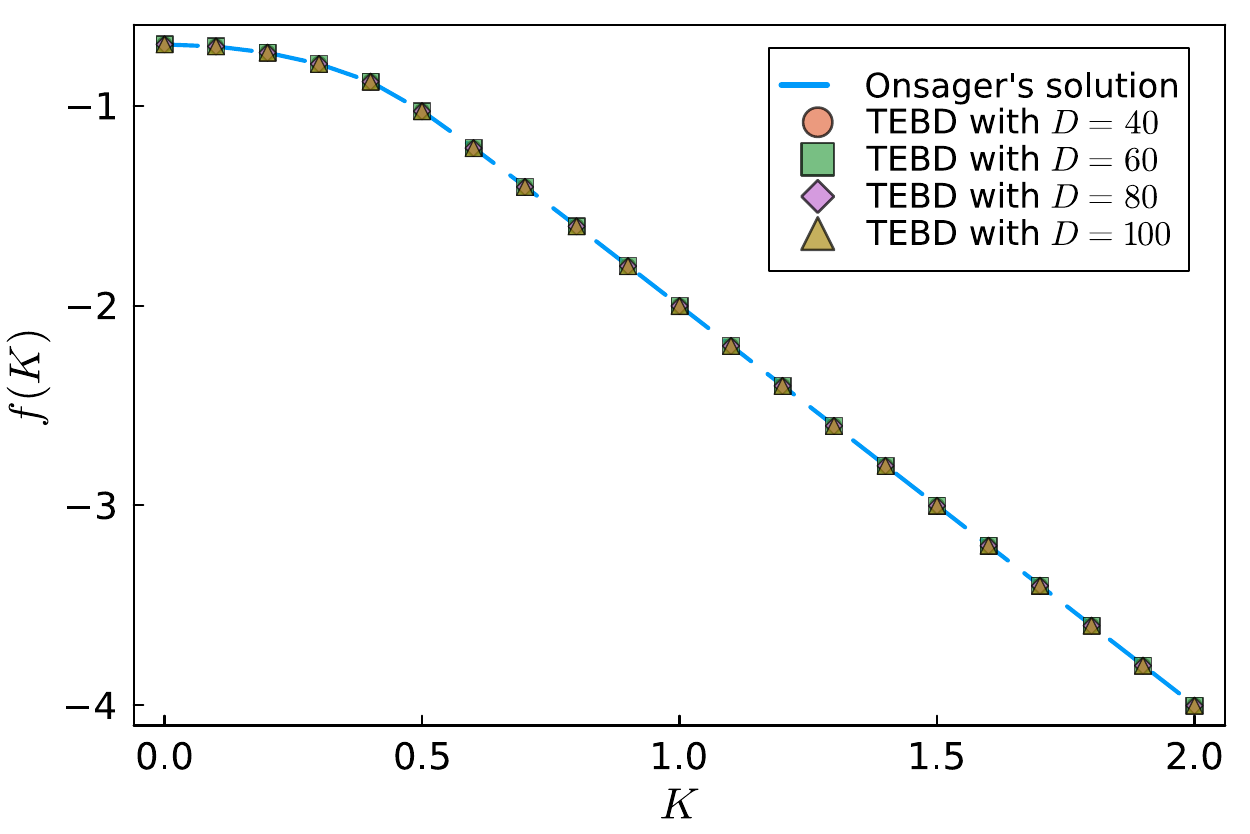}
    \caption{Bulk free-energy density of two-dimensional Ising model. The dashed line: Onsager's solution; symbols: TEBD simulation with bond dimensions $D = 40,\ 60,\ 80,\ 100$ for a $20\times20$ torus. All markers lie exactly on the analytical curve.}
    \label{fig: TEBD benchmark}
\end{figure}

We simulated the free-energy density of the 2D Ising model on a $20 \times 20$ torus over $0 \leq \beta J \leq 2$. To evaluate the effect of the SVD bond-dimension cutoff, we compare $D = 40,\, 60,\, 80,\, 100$. Fig. \ref{fig: TEBD benchmark} shows that all TEBD data collapse onto Onsager's solution within plot resolution, indicating that $D=40$ already saturates the accuracy required for our RBIM computations.

\section*{The efficiency of postselection}

\textbf{Postselecting on the maximal $|\Delta F_i|$ among $r$ independent copies. }
As introduced in the main text, one strategy is to prepare $r$ independent copies of the toric code, each with code distance $d$, and postselect on the block with the largest free-energy difference magnitude $|\Delta F_i|$. We claim that the logical failure probability under such postselection is
\begin{equation}
    \label{eq: LDP SM}
    p_f^{{(r)}}(d)\sim \exp\{-\min_{s\ge0}[rI(s)+s]d\}~.
\end{equation}
We derive this form via maximum order statistics, which provides a baseline for analyzing postselection efficiency at fixed qubit overhead. For a distance-$d$ code, the left-tail PDF of the free-energy difference $\Delta F$ obeys
\begin{equation}
    P_d(\Delta F)\sim\exp(-I(s)d), \qquad I(\overline{s})=0, \quad I(s)+s=I(-s),
\end{equation}
for $|s|\leq \overline{s}$. The cumulative distribution function (CDF) of $|\Delta F|$ is
\begin{equation}
    \mathcal{F}_d(M)\equiv\int_{-M}^{M}P(\Delta F)\dd \Delta F~, \qquad M\ge 0~.
\end{equation}

From order statistics, the PDF of the maximum $|\Delta F_i|$ among $r$ copies is
\begin{equation}
    P_d(|\Delta F_i|_{\max}=M)=r\,\mathcal{F}_d(M)^{r-1}\,[P_d(\Delta F=M)+P_d(\Delta F=-M)]~.
    \label{eq: maximum order statistic integrand}
\end{equation}
Because of the symmetry $P(\Delta F)=P(-\Delta F)e^{\beta\Delta F}$, the logical failure probability can be written as
\begin{equation}
    p_f=\int_{-\infty}^{0} P(\Delta F)\dd \Delta F =\int_{-\infty}^\infty\frac{P(\Delta F)}{1+e^{|\Delta F|}}\dd \Delta F.
\end{equation}
Therefore, after postselection, the logical failure probability is
\begin{equation}
    \label{eq: p_log with n copies}
    p_{f}^{(r)}(d) = \int_0^\infty \frac{P_d(|\Delta F_i|_{\max}=M)}{1+e^{M}} \dd M~.
\end{equation}
Since $P(\Delta F)$ obeys large deviation scaling, for $M\leq\overline{s}d$, the CDF behaves as
\begin{equation}
    \mathcal{F}_d(M)=\int_{|\Delta F|\le M} P(\Delta F)\dd \Delta F\sim \exp\!\left[-I\!\left(\tfrac{M}{d}\right)d\right],
\end{equation}
and Eq.~\eqref{eq: maximum order statistic integrand} gives
\begin{equation}
    P_d(|\Delta F_i|_{\max}=M)\sim \mathcal{F}_d(M)^{r-1}P_d(\Delta F=M)\sim \exp\!\left[-rI\!\left(\tfrac{M}{d}\right)d\right].
\end{equation}
Substituting into Eq.~\eqref{eq: p_log with n copies} yields
\begin{equation}
    p_{f}^{(r)}(d) \sim \int_{\mathbb{R}^{+}}\exp\!\left(-rI\!\left(\tfrac{M}{d}\right)d - M\right)\dd M
    \sim\exp\{-\min_{s\ge0}[rI(s)+s]d\}~,
    \label{eq: MOS logical error rate}
\end{equation}
as stated in the main text.

\textbf{Postselection efficiency for splitting into four subcodes. }
By convexity of $I(s)$, we showed in the main text that two distance-$d$ codes scale equivalently to a single distance-$2d$ code. Consequently, splitting a distance-$d$ toric code into two distance-$(d/\sqrt{2})$ codes suppresses logical errors as
\begin{equation}
    p_f^{(2)}(d/\sqrt{2})\simeq e^{-\sqrt{2}I(0)d}\ll e^{-I(0)d}\simeq p_f(d).
\end{equation}
Because $\sqrt{2}$ is irrational, it is not feasible to directly simulate both distance-$d$ and distance-$(d/\sqrt{2})$ codes. Instead, we compared $p_{f}(2n)$, $p^{(2)}_{f}(n)$, and $p^{(4)}_{f}(n)$, and observed numerically that $p^{(4)}_{f}(n)$ decays exponentially faster in $n$. We now provide the mathematical explanation.

Selecting the maximal $|\Delta F_i|$ among four distance-$n$ patches gives
\begin{equation}
    p_{f}^{(4)}(n)\simeq \exp\{-\min_{s\ge0}[4I(s)+s]d\}.
\end{equation}
Since $I(s)+s=I(-s)$ implies $I'(0)=-1/2$, convexity guarantees that $4I(s)+s$ has a unique global minimum at some $s_m>0$. As $s=0$ is the minimum for $2I(s)+s$, one finds
\begin{equation}
    \min_{s\ge0}[4I(s)+s]=4 I(s_m)+s_m\ge2I(s_m)+s_m>\min_{s\ge0}[2I(s)+s]=2I(0).
\end{equation}

Thus, splitting into four always provides a positive gain, demonstrating the advantage of postselection under a fixed qubit budget. This improvement, however, does not extend to arbitrary $r$. Since $\min_{s\ge0}[rI(s)+s]\le\overline{s}$, when $r>b^2$ one obtains
\begin{equation}
    p^{(r)}_{f}(\frac{d}{\sqrt{r}})\simeq \exp\!\left\{-\tfrac{\min_{s\ge0}[rI(s)+s]}{\sqrt{r}}\,d\right\}\ge e^{-\overline{s}d/\sqrt{r}}>e^{-I(0)d}\simeq p_{f}(d).
\end{equation}

Hence, the efficiency gain of code-splitting eventually diminishes as $r$ increases. 

\section*{Postselection for general stabilizer codes}
\textbf{Marginal decoding. }In the main text we have argued that for a toric code with independent $X$ and $Z$ errors and an optimal decoder one can define the free-energy differences $\Delta F$, and most logical errors occur due to $\Delta F$ near zero. 
However, such arguments might be flawed when (i) $X$ and $Z$ errors are correlated, for instance, depolarizing noise; (ii) for more general qubit stabilizer codes. Here we try to generalize our previous discussion into general error models and more general codes.

Consider a general $[[n,k,d]]$ qubit stabilizer code with logical operators $\{L_i^X, L_i^Z\}$ on the $i^{\text{th}}$ logical qubit ($1\le i \le k$). Under maximum-likelihood decoding (MLD), one constructs a canonical error chain $\tilde{E}$ consistent with the measured syndrome and evaluates $P([\tilde{E}\mathcal{P}_l])$ (where $[\cdot]$ denotes an error class modulo stabilizers) for all $4^k$ logical Pauli operators $\mathcal{P}_l$ generated by $\{L_i^X,L_i^Z\}$. Decoding then selects $\arg\max P([\tilde{E}\mathcal{P}_l])$.

Let $\Omega$ denote the probability space of all physical errors. For any error $E\in \Omega$, we may define 
\begin{equation}
    \Delta F_l(E) \equiv \log \frac{P([E])}{P([E\mathcal{P}_l])}~.
\end{equation}

For large $k$, evaluating all $(4^k-1)$ quantities is infeasible. An efficient alternative is marginal decoding~\cite{PhysRevResearch.7.013040}, which assigns each logical qubit to the most likely single-qubit Pauli class by marginalizing over errors on the other $(k-1)$ qubits. This procedure effectively projects multi-qubit errors onto single-qubit marginals. More precisely, the marginal decoder defines the probability of a single-qubit logical error on the $i^{\text{th}}$ qubit by summing over all logical errors on the other $(k-1)$ qubits:  
\begin{equation}
    P_m([E L_{i}^{\alpha}])=\sum_{l=1}^{4^{k-1}} P([E L_{i}^{\alpha} \mathcal{P}_l^{(-i)}]),  
    \label{eq: marginal probability}
\end{equation}
where $L_{i}^{\alpha}\in\{L_i^I,L_i^X,L_i^Y,L_i^Z\}$ and $\mathcal{P}_l^{(-i)}$ are generated by $\{L_j^X,L_j^Z\mid j\neq i\}$. 

In the decoding process, the decoder determines some canonical error chain $\tilde{E}$ and then seeks the operator $L_i^{\alpha_{\max}}$ maximizing $P_m([\tilde{E} L_{i}^{\alpha}])$, yielding the final correction as $E=\tilde{E}\prod_i L_i^{\alpha_{\max}}$.

For each nontrivial $L_{i}^{\alpha}$ ($\alpha=X, Y, Z$), given the actual error $E\in\Omega$, we define a free-energy difference
\begin{equation}
    \Delta F_{i, \alpha}\equiv \log\frac{P_m([E])}{P_m([E L_{i}^{\alpha}])}.
\end{equation}
There are $3k$ such quantities (corresponding to $L_i^X,L_i^Y,L_i^Z$). According to the marginal probability definition in Eq.~\eqref{eq: marginal probability}, all error chains $E'\in[EP_{l}^{(-i)}]$ share the same marginal probability, i.e., $P_m([E'L_{i}^{\alpha}])=P_m([EL_{i}^{\alpha}])$. Moreover, the error chains $E\in[EP_l^{(-i)}]$ yield $\Delta F_{i, \alpha}$ values opposite to those of the chains $E\in[EL_{i}^{\alpha}P_l^{(-i)}]$. By construction, $\Delta F_{i, \alpha}$ therefore satisfies the same constraint as in the $k=1$ case.

\begin{equation}\label{eq:marginal-nishimori}
    P(\Delta F_{i, \alpha})=P(-\Delta F_{i, \alpha})e^{\Delta F_{i, \alpha}},
\end{equation}
and are expected to satisfy a large deviation principle (LDP) of the form
\begin{equation}
\label{eq: LDP}
    P(\Delta F_{i, \alpha}=sd)\simeq e^{-d I_{i, \alpha}(s)},
\end{equation}

with $I_{i, \alpha}(s)$ convex, monotonically decreasing for $s\le \overline{s}_{i, \alpha}$, obeying $I_{i, \alpha}(s)+s=I_{i, \alpha}(-s)$, and vanishing as $I_{i, \alpha}(\overline{s}_{i, \alpha})=0$. These properties imply
\begin{equation}
    \overline{s}_{i, \alpha}=I_{i, \alpha}(\overline{s}_{i, \alpha})+I_{i, \alpha}(-\overline{s}_{i, \alpha})\ge 2I_{i, \alpha}(0).
\end{equation}

Notably, it has been revealed that the marginal decoding is also optimal when the physical error rate is below the threshold and the code distance $d$ is sufficiently large \cite{PhysRevA.105.052446}, providing a generally feasible approach to decode a stabilizer code with multiple logical qubits.

\textbf{Efficiency of postselection. }In either MLD or marginal decoding, the problem reduces to evaluating $\mathcal{M}$ free-energy differences $\Delta F_i$, with $\mathcal{M}=3k$ for marginal decoding and $\mathcal{M}=4^k-1$ for MLD. Decoding succeeds iff all $\Delta F_i\ge0$. For scalability, $\mathcal{M}$ must grow subexponentially with $d$, i.e.
\begin{equation}
    \label{eq: subexp growth of M}
    \lim_{d\to\infty}\mathcal{M}e^{-cd}=0,\qquad \forall c>0.
\end{equation}

Let $\Omega$ be the space of physical errors, and $\mathcal{W}_i=\{E\in\Omega|\Delta F_i(E)<0\}$. The set of logical failures is $\mathcal{W}=\bigcup_i\mathcal{W}_i$, with probability bounded as
\begin{equation}
    \max_i P(\mathcal{W}_i)\ \le\ P(\mathcal{W})\ \le\ \mathcal{M}\,\max_i P(\mathcal{W}_i).
\end{equation}
thus the dominant contribution is controlled by the largest $P(\mathcal{W}_i)$.  

A natural postselection strategy is to discard all errors with $|\Delta F_i|<M_i$, choosing thresholds $M_i$ such that
\begin{equation}
    P(\Delta F_i=-M_i)=P(\Delta F_j=-M_j),
\end{equation}
Let $\mathcal{A}_i=\{E\in\Omega|M_i>|\Delta F_i(E)|\}$ and $\mathcal{A}=\bigcup_i \mathcal{A}_i$ be the abort set. Then
\begin{equation}
    \max_i P(\mathcal{A}_i)\ \le\ P(\mathcal{A})\ \le\ \mathcal{M}\,\max_i P(\mathcal{A}_i).
\end{equation}
Since each $P(\Delta F_i)$ has the same large deviation properties as in the surface-code setting, we propose discarding syndromes with $|\Delta F_i|<s^{*}_i d$, where the $s^{*}_i$ are chosen such that $I_i(-s^{*}_i)$ is equal across $i$ and $s^{*}_i\le \overline{s}_i$ to ensure finite acceptance probability. The logical failure rate after postselection scales as
\begin{equation}
    p_f\ \simeq\ \sum_i e^{-I_i(-s^{*}_i)d}\ \ge\ e^{-\min \overline{s}_i\,d}.
\end{equation}
The second inequality holds since because all $I_i(-s^*_i)$ are equal and $k$ is subexponential in code distance $d$. By contrast, without postselection one has
\begin{equation}
    p_f\ \simeq\ e^{-\min I_i(0)\,d}.
\end{equation}
Hence the maximum possible gain from postselection is quantified by
\begin{equation}
    b\equiv \frac{\min I_i(-\overline{s}_i)}{\min I_i(0)}\ \ge\ \frac{2\min I_i(0)}{\min I_i(0)}=2.
\end{equation}

Thus, for any stabilizer code, postselection can improve the effective distance by at least a factor of two, with the precise value of $b$ determined by the large deviation functions $I_i(s)$.

\section*{Verifications of large deviation scaling}
\textbf{MWPM decoding. }As discussed in the main text, both MLD decoders and marginal decoders that obey large deviation scaling within certain regimes provide a scalable accuracy gain with $b \ge 2$. However, evaluating the exact probability of an error class becomes impractical under circuit-level noise, where measurement and CNOT gate errors must be taken into account. In practice, minimum-weight perfect matching (MWPM) decoding is typically employed to efficiently handle circuit-level noise. In this setting, the complementary gap $G$ serves as a key indicator of code performance~\cite{Gidney2025}. For every error configuration $E$, let 
$\mathcal{C}_{\text{corr}}(E)$ and $\mathcal{C}_{\text{inc}}(E)$
denote the sets of error chains in the correct and incorrect homology
classes, respectively. Define 
\begin{subequations}
\begin{align}
w_{\text{success}}(E) &= \min_{\mathcal{E} \in \mathcal{C}_{\text{corr}}(E)} -\log P(\mathcal{E}), \\
w_{\text{fail}}(E) &= \min_{\mathcal{E} \in \mathcal{C}_{\text{inc}}(E)} -\log P(\mathcal{E}) .
\label{Eq:w_success_fail_gap}
\end{align}
\end{subequations}
A MWPM decoder always selects the class containing the matching with the minimum weight; in particular, the correct class is chosen if $w_{\rm success} < w_{\rm fail}$. Analogous to the definition of $\Delta F$, the complementary gap $G$ \cite{Gidney2025, PhysRevA.89.022326, PRXQuantum.5.010302} in MWPM is defined as 
\begin{equation}
    G = w_{\text{fail}} - w_{\text{success}},
\label{Eq:G_def}
\end{equation}
which plays the same role for MWPM as the free-energy difference $\Delta F$ does in MLD, but is computationally more tractable. From a statistical-physics perspective, MWPM decodes at zero temperature rather than along the Nishimori line. Using Stim~\cite{gidney2021stim} and PyMatching~\cite{higgott_pymatching_2022}, we numerically investigated the distribution $P(G)$ for the rotated surface code under both the code-capacity and circuit-level noise models. 
In the rotated surface code~\cite{PhysRevA.76.012305,Horsman_2012}, boundaries are classified as $X$-type or $Z$-type, depending on whether $X$- or $Z$-syndrome checks are measured along that boundary. Logical operators are Pauli strings that connect opposite boundaries of the same type, with $X_L$ connecting the two $X$-type boundaries and $Z_L$ connecting the two $Z$-type boundaries.

\begin{figure}[t]
    \centering
    \includegraphics[width=0.8\linewidth]{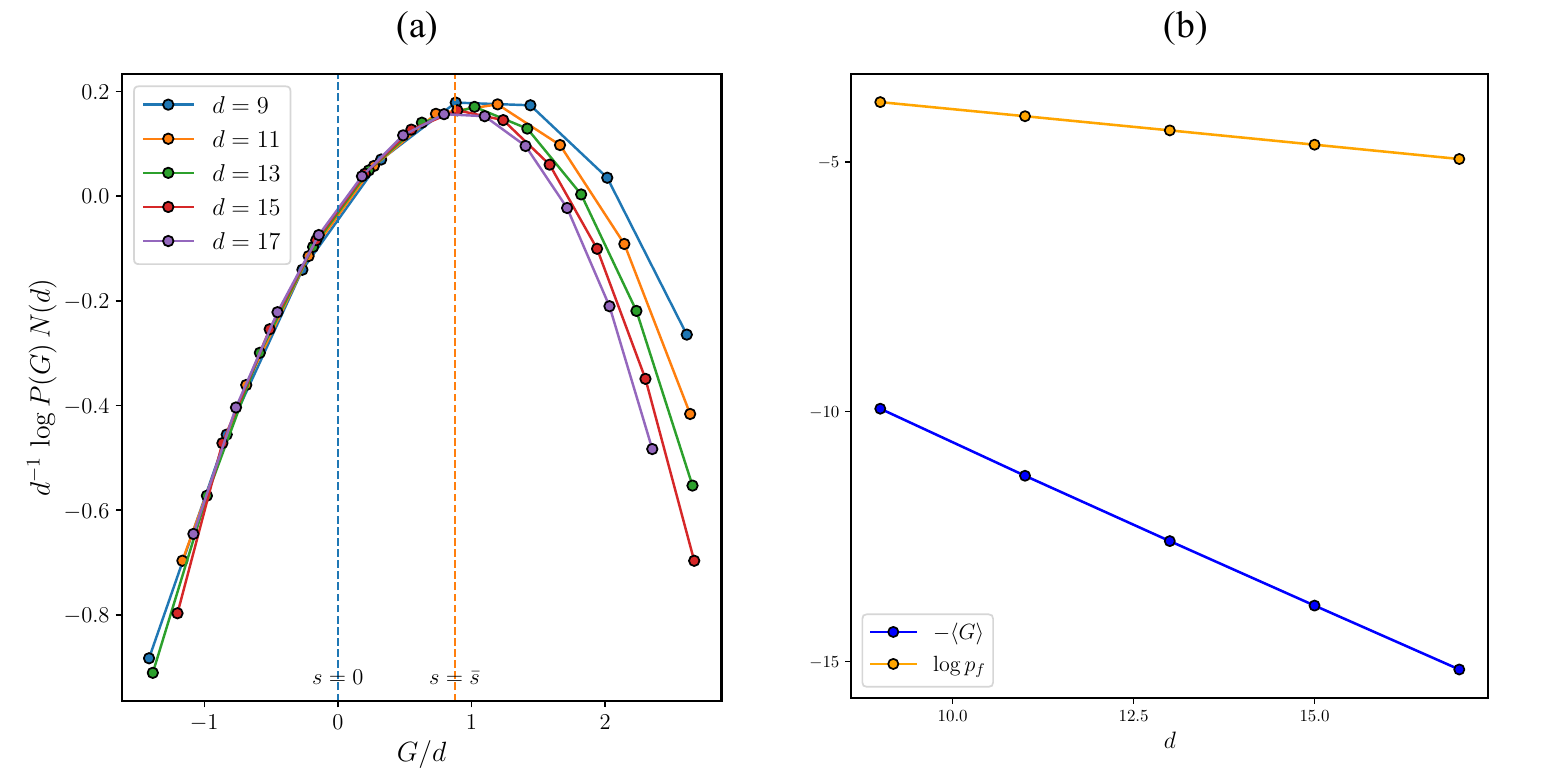}
    \caption{Distribution of $P(G)$ in MWPM decoding of rotated surface code under $p=0.06$ bit-flip noise and perfect stabilizer measurements. (a) Verification of large deviation scaling in two-dimensional MWPM. Still, the scaling holds for $|G|\le\langle G\rangle$. (b) Scaling of $\log p_f$ and $-\langle G\rangle$ with code distance $d$. Fitting from $\log p_f$ gives $I(0)=0.142$, while $\overline{s}=0.652$ is determined from $\langle G\rangle$ scaling. The factor $c$ is fitted as $0.75$, which fixes the maximal gain in effective code distance as $b=3.44$. }
    \label{fig: MWPM 2D}
\end{figure}

\begin{figure}[t]
    \centering
    \includegraphics[width=1\linewidth]{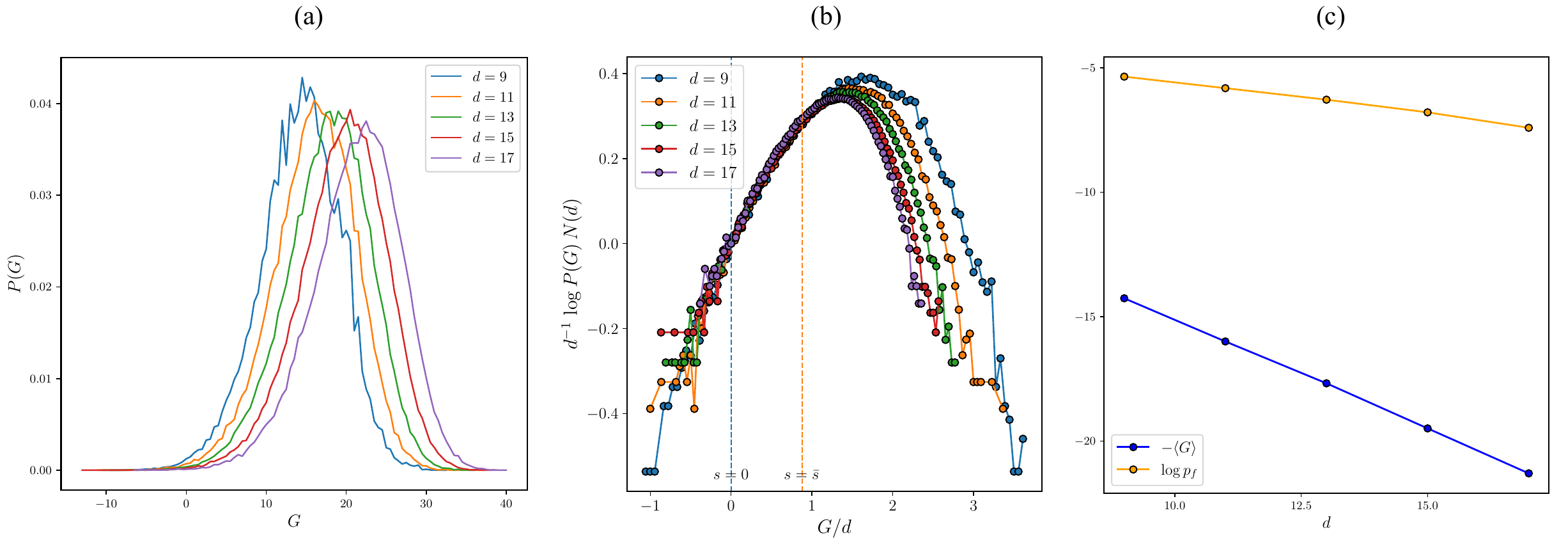}
    \caption{Probability distribution $P(G)$ in MWPM decoding of the rotated surface code under circuit-level noise, with $p=0.005$ error rate for CNOT gates, depolarizing noise, and measurement errors. Syndromes are extracted with $d$ rounds of stabilizer measurements. The complementary gap $G$ is defined as the difference in MWPM weights between competing logical error classes, and can be computed separately for logical $X_L$ (using the $Z$-check graph) and logical $Z_L$ (using the $X$-check graph). Because the noise model is symmetric between $X$ and $Z$, the resulting distributions $P(G)$ are similar, so only the $Z$-check case is shown here. 
(a) Unscaled distribution of gaps $G$ for different code distances $d$. 
(b) Verification of large deviation scaling in three-dimensional MWPM. 
(c) Scaling of the logical failure rate $\log p_f$ and the average gap $-\langle G\rangle$ with code distance $d$. A fit from $\log p_f$ yields $I(0)=0.254$, while $\overline{s}=0.878$ is obtained from the scaling of $\langle G\rangle$. The prefactor $c$ is fitted as $0.83$, giving $b=2.87$.}

    \label{fig: MWPM 3D}
\end{figure}

As shown in Fig.~\ref{fig: MWPM 2D} and Fig.~\ref{fig: MWPM 3D}, the gap distributions $P(G)$ satisfy large deviation scaling within the range $|G|\le\langle G\rangle$ as $P(G=sd)\sim e^{-I(s)d}$, and we continue to presume the monotonicity and convexity of $I(s)$.

However, the distribution of the complementary gap breaks the symmetry $P(\Delta F)=P(-\Delta F)e^{\Delta F}$, and instead appears to obey~\cite{Gidney2025}
\begin{equation}
    P(G)=P(-G)e^{cG},
    \label{eq: modified constraint}
\end{equation}
with $c$ fitted from the distribution $P(G)$. This enables us to estimate $I(-\overline{s})$ using the scaling of $\langle G\rangle$. Furthermore, the maximum possible gain from postselection is still bounded by
\begin{equation}
    b\equiv\frac{I(-\overline{s})}{I(0)}=\frac{I(-\overline{s})+I(\overline{s})}{I(0)}\ge 2.
\end{equation}

Thus, replacing the MLD decoder with the MWPM decoder does not change the lower bound of $b$, implying a finite improvement in MWPM decoding via postselection. Numerically, decoding under circuit-level noise with $p=0.005$ gives $b\approx2.87$, while the gap distribution of the rotated surface code with perfect stabilizer measurement at bit-flip error rate $p=0.06$ yields $b\approx 3.44$, in contrast to the $b\approx3$ result in MLD.

\textbf{Generalized Shor code. }We next verify large deviation scaling in a nontopological stabilizer code: a family of concatenated generalized Shor (GS) codes~\cite{Shor1995}, obtained by alternatively concatenating a two-qubit repetition code in the $X$ basis (with stabilizer group $\langle ZZ \rangle$) and a two-qubit repetition code in the $Z$ basis (stabilizer group $\langle XX\rangle$). A GS code with $t$ concatenation layers can be visualized as a depth $t$ binary tree in which a single logical qubit fed into the root is encoded into $2^t$ physical qubits on the leaves~\cite{Yadavalli2025,PhysRevResearch.7.023086}. The parameters of this code are $[[n=2^t, k=1, d=2^{\lfloor t/2\rfloor}]]$. 

\begin{figure}[bt]
\includegraphics[width=\linewidth]{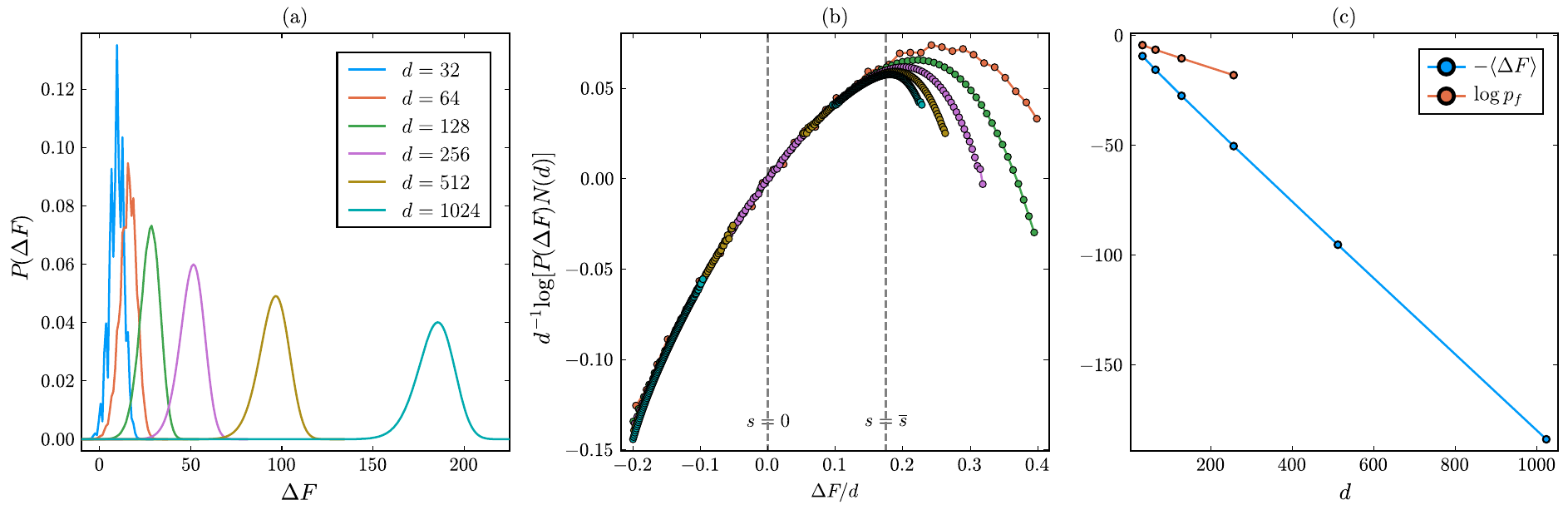}
\caption{Distribution of $\Delta F$ in the concatenated GS code, under bit flip errors at rate $p=0.1$. Data were obtained via population dynamics for a total of $8 \times 10^8$ samples, and binned by integers. (a) Unscaled distribution. (b) Scaled distribution, with the smallest system size ($d=32$) omitted. Gray dashed lines mark $s=0$ and $s=\overline{s}$, the latter obtained via a linear fit to $\langle \Delta F\rangle$ using $d \geq 256$. (c) $-\langle \Delta F \rangle$ (blue) and $\log p_f$ (orange). We were unable to obtain an accurate estimate of $\log p_f$ for $d\geq 512$, due to a lack of signal near $\Delta F = 0$ at these very low logical error rates.\label{fig:gs}}
\end{figure}

We now consider applying bit flip noise to the leaves of this tree. The tree structure of the encoding circuit enables an efficient maximum likelihood decoder. However, rather than corresponding to the free energy cost of a system-spanning domain wall as it does in the toric code, $\Delta F$ now corresponds to the effective field at the root of the tree induced by a certain boundary condition on the leaves.

Fig. \ref{fig:gs} shows the  $P(\Delta F)$ under bit flip errors at rate $p=0.1 < p_{c,X} \approx 0.148$, for even depths up to $t=20$. Following the method described in Ref.~\cite{PhysRevResearch.7.023086}, we performed 40 independent runs of population dynamics with 2 populations each of size $10^7$, for a total of $8 \times 10^8$ correlated samples. For $t=18$ and $t=20$ ($d=512$ and $d=1024$), this sample size was sufficient to resolve the peak of the distribution (and thus $\langle \Delta F \rangle$), but not to access the behavior near $\Delta F = 0$, as the logical failure probability decays doubly exponentially with $t$. Nevertheless, with the accessible data, we observe a collapse to the large deviation scaling form in the left tail (Fig.~\ref{fig:gs}(b)). As for the toric code in the main text, we also estimate $\overline{s}$ and $I(0)$ from linear fits to $\langle \Delta F \rangle = F_0 + \overline{s}d$ and $\log p_f = L_0 - I(0) d$, respectively. Both are sensitive to the range of the linear fit, shifting to smaller values if the fitting interval is moved to larger $d$. Thus, $\overline{s} = 0.174$ (inferred from a fit using $d=256,512,1024$) and $I(0) = 0.060$ (inferred from a fit using $d=64,128,256$) should both be treated as upper bounds. Despite this uncertainty, the data confirm the ratio $b \geq 2$, as expected from convexity.

\section*{Observation of $b\approx 3$ in two-dimensional RBIM}
\textbf{Partially heralded error and bond-diluted RBIM.}  
In the main text we quantify the gain in effective code distance by $b \equiv I(-\overline{s})/I(0)$. In the extreme case where all Pauli errors are heralded, logical failures occur only when bonds percolate \cite{PhysRevLett.102.200501}. Since such percolation events form an exponentially small fraction of error chains, aborting them yields perfect logical fidelity ($b=\infty$). A natural next step is to ask: what is the value of $b$ when only a fraction of Pauli errors are heralded \cite{Wu_2022, chang2024surfacecodeimperfecterasure}?  

Suppose each qubit undergoes a \emph{partially heralded} bit-flip channel: with probability $p r_e$ the error is flagged as an erasure, with probability $p(1-r_e)$ the error is unflagged, and with probability $1-p$ no error occurs. Thus each qubit experiences a mixture of two channels:  

\noindent(i) \textbf{Heralded branch ($\varepsilon_e=1$)} — with probability $p r_e$:  
\begin{equation}
\mathcal{E}_{\varepsilon_e=1}(\rho)=\tfrac14\!\left(I\rho I+X\rho X+Y\rho Y+Z\rho Z\right),
\label{eq:flag_channel}
\end{equation}
(ii) \textbf{Unheralded branch ($\varepsilon_e=0$)} — with probability $p(1-r_e)$:  
\begin{equation}
\mathcal{E}_{\varepsilon_e=0}(\rho)=(1-p(1-r_e))\,\rho + p(1-r_e)\,X\rho X,
\label{eq:unflag_channel}
\end{equation}
where the decoder is unaware of the flip.  

Including the no-error component $(1-p)\rho$, the full channel is  
\begin{equation}
\mathcal{E}(\rho)=(1-p)\rho + p r_e\,\mathcal{E}_{\varepsilon_e=1}(\rho) + p(1-r_e)\,\mathcal{E}_{\varepsilon_e=0}(\rho),
\label{eq:ph_channel_total}
\end{equation}
with $\varepsilon_e\in\{0,1\}$ revealed to the decoder. Conditional on $\varepsilon_e=0$, the effective flip probability is  
\[
p_{\mathrm{eff}}=\frac{p(1-r_e)}{1-p r_e},\qquad
K_e=\tfrac12\log\!\Bigl(\tfrac{1-p_{\mathrm{eff}}}{p_{\mathrm{eff}}}\Bigr).
\]
Hence the bond distribution in the RBIM is
\[
\boxed{P(J_e)=p r_e\,\delta(J_e)+(1-p)\,\delta(J_e-K_e)+p(1-r_e)\,\delta(J_e+K_e)},
\]
with Nishimori relation $p_{\mathrm{eff}}=1/(1+e^{2K_e})$. Non-erased edges carry $J_e=K_e$, while flagged edges correspond to $J_e=0$ and are deleted from the dual Ising graph. Thus the noise maps to an RBIM with bond dilution at density $p r_e$.  

Two limits illustrate the model:  
(i) $r_e=0$ gives $p_{\mathrm{eff}}=p$, i.e. the standard RBIM on the Nishimori line.  
(ii) $r_e\to1$ (with $p$ finite) yields vanishing unflagged flips and diverging couplings ($J_e\to\infty$); decoding is then determined by percolation of erased bonds ($J_e=0$) at density $p$.  

\begin{figure}[t]
    \centering
    \includegraphics[width=0.9\linewidth]{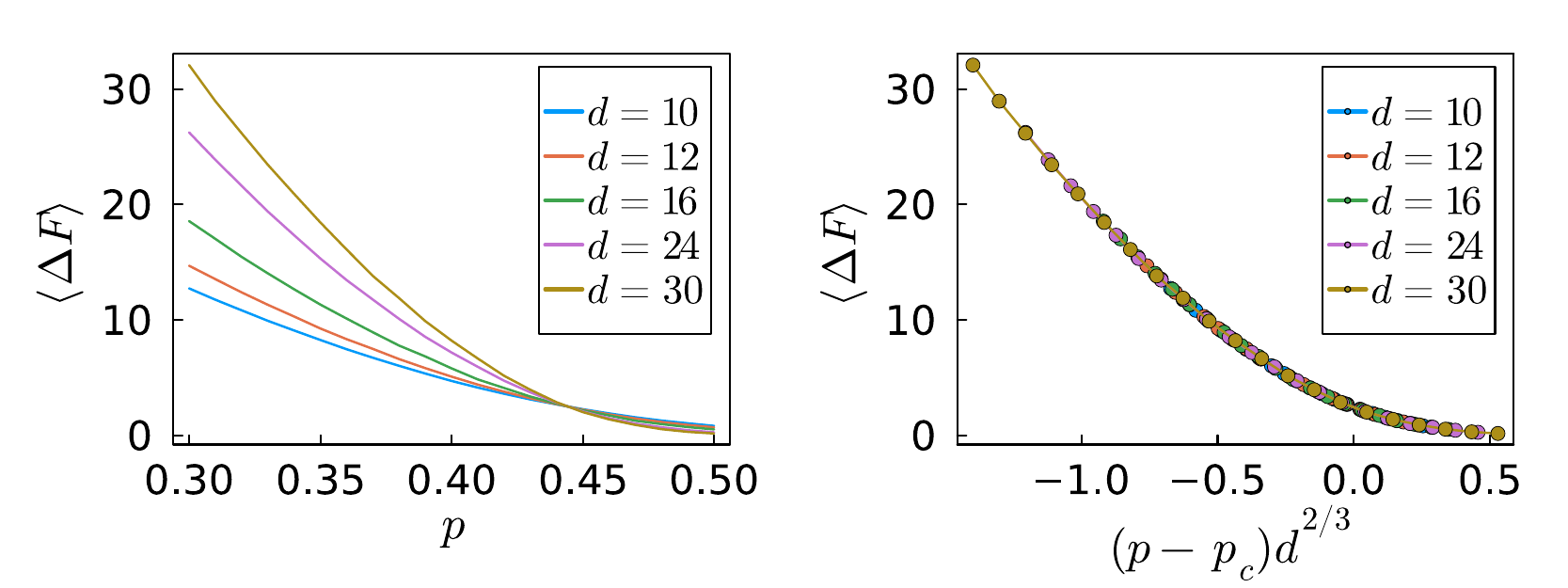}
    \caption{Finite-size scaling of the average free-energy difference $\langle\Delta F\rangle$ in the random-bond Ising model with partial heralding probability $r_e=0.98$.  
\textbf{(a)} $\langle\Delta F\rangle$ versus physical error rate $p$ for code distances $d=10,12,16,24,30$; the curves intersect at the critical point $p_c\approx0.445$.  
\textbf{(b)} Data collapse near the transition when plotted against the scaling variable $(p-p_c)\,d^{1/\nu}$ with $\nu=\tfrac{3}{2}$. The agreement with the un-heralded ($r_e=0$) case confirms that all RBIMs on the Nishimori line belong to the same universality class, except for the point $r_e=1$, which is unstable towards partially heralding. 
}
    \label{fig:universality class}
\end{figure}

Importantly, changing the bond distribution does not affect the universality of the transition. Near the critical point $p_c(r_e)$, the domain-wall free energy scales as  
\begin{equation}
\Delta F(p,d;r_e)=f\!\left(\frac{p-p_c(r_e)}{d^{1/\nu}}\right),\qquad \nu\simeq1.5,
\end{equation}
with $d$ the lattice size. Fig. \ref{fig:universality class} shows that data for $r_e=0.98$ collapse with $\nu=3/2$, matching the undiluted case ($r_e=0$). Thus all partially heralded RBIMs with $0\le r_e<1$ belong to the same universality class \cite{Picco_2006}. In contrast, the percolation fixed point at $r_e=1$ and $p=0.5$ exhibits $\nu_{\mathrm{perc}}=4/3$, marking a distinct universality class \cite{smirnov2001criticalexponentstwodimensionalpercolation}.  

\textbf{Observation of $b\approx3$ in bond-diluted two-dimensional RBIM.}  
We now quantify the distance gain $b$ in the diluted RBIM using the free-energy difference $\Delta F$. Recall that $b=\overline{s}/I(0)$, where $I(0)$ is extracted from the scaling of $\log p_f$ and $\overline{s}$ from the linear growth of $\langle \Delta F\rangle$. In the limiting case $r_e=1$ the model reduces to percolation of erased bonds; aborting the exponentially rare percolating instances forces the logical failure rate to vanish faster than any fixed-rate large deviation cost, so that $b\to\infty$ at fixed code distance. Consequently, at small $d$ one expects $b$ to increase as $r_e$ is tuned upward toward $1$.

\begin{figure}[t]
    \centering
    \includegraphics[width=0.85\linewidth]{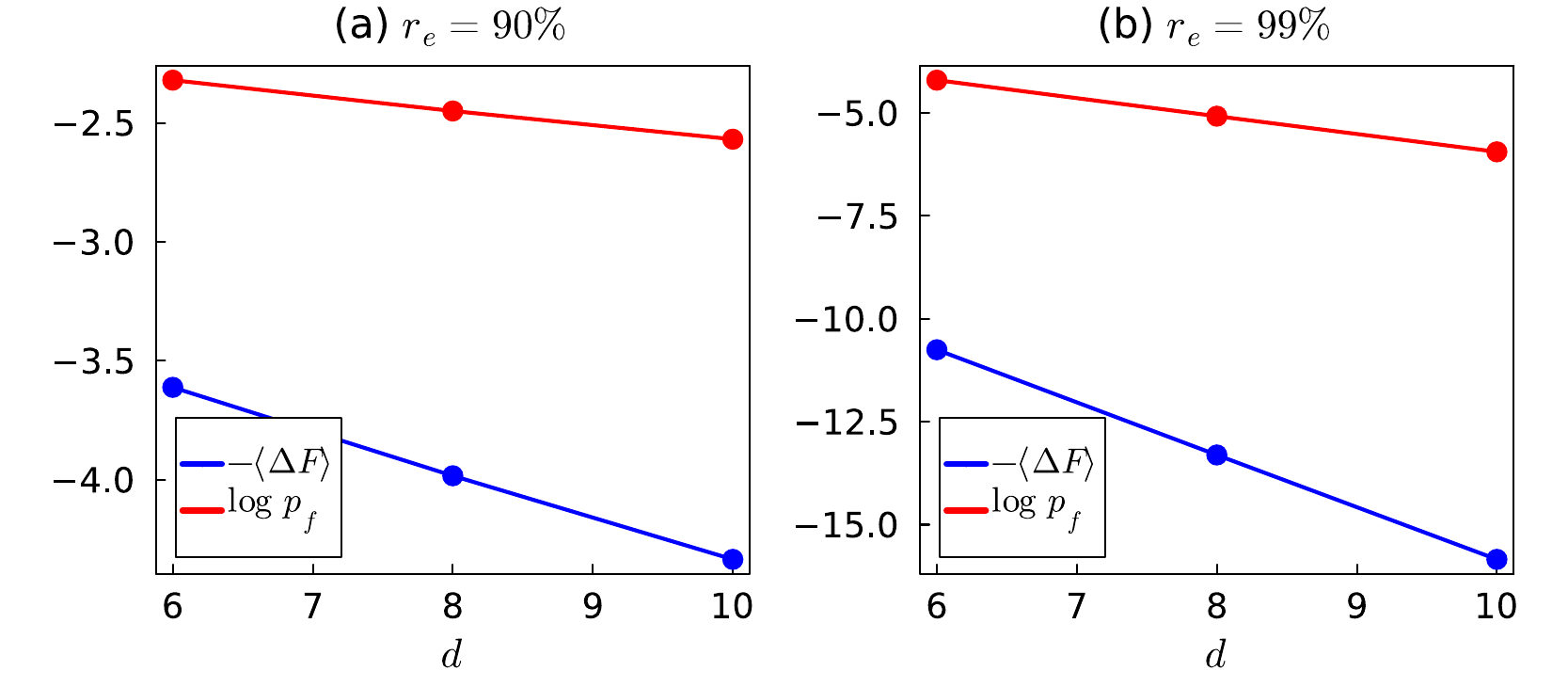}
    \caption{Numerical evalution of $b$ in bond-diluted $2D$ RBIM with $d=6, 8, 10$ and $p=0.3$. \textbf{(a)} $r_e=90\% $ RBIM. Fitting from $\langle\Delta F\rangle$ gives $\overline{s}=0.18$, while $I(0)=0.062$ is evaluated from $\log p_f$ scaling, showing $b=2.91$. \textbf{(b)} $r_e=99\% $ RBIM. $\overline{s}=1.27$ and $I(0)=0.435$ are evaluated from $\langle\Delta F\rangle$ and $\log p_f$ scalings, showing $b=2.92$. 
    }
    \label{fig: partially heralded}
\end{figure}

Our finite-size simulations, however, reveal a different picture. Fig. \ref{fig: partially heralded} shows that even when $r_e$ is very close to one ($90\%$ and $99\%$), the fitted values remain around $b\approx 3$. Specifically, we find $b=2.91$ at $r_e=90\%$ and $b=2.92$ at $r_e=99\%$ using data from $d=6,8,10$. This suggests that, within the accessible sizes, increasing the heralded rate does not significantly change the value of $b$, and the system continues to behave similarly to the RBIM without dilution.

A heuristic rationale for $b\approx 3$ can be obtained by extrapolating the Tracy--Widom distribution near $\Delta F\simeq\langle\Delta F\rangle$. In this regime one expects  
\[
I(s)=c_0\,(\overline{s}-s)^{3/2}, \qquad s\le \overline{s}.
\]  
Using the symmetry relation $I(s)+s=I(-s)$ to evaluate $I'(0)$ gives $c_0=1/\sqrt{9\overline{s}}$ and hence $I(0)=\overline{s}/3$, implying $b=3$. However, the Tracy--Widom form is rigorously justified only for fluctuations $|\Delta F-\langle\Delta F\rangle|\sim d^{1/3}$, corresponding to $|s-\overline{s}|\sim d^{-2/3}$. Outside this narrow window the functional form of $I(s)$ is unknown, so at present we cannot claim a derivation of $b=3$, only a suggestive consistency.

\section*{Analysis of the scaled cumulant generating function $\lambda(k)$}
\textbf{G\"artner--Ellis theorem.}
In the main text and previous sections, we numerically demonstrated the validity of the large-deviation principle (LDP) for several code families, including the concatenated generalized Shor code and the toric code under both code-capacity and circuit-level noise models. Moreover, in topological codes, the distribution of $\Delta F$ is naturally related to free-energy fluctuations of disordered domain walls and DPRM, which are themselves generally expected to obey large-deviation behavior. Nevertheless, it remains unclear why an LDP should generically apply to arbitrary quantum error-correcting codes. Although a fully rigorous proof of the LDP is expected to be highly nontrivial, we now provide a heuristic argument based on the G\"artner--Ellis theorem, which relates the validity of the LDP to the regularity properties of the scaled cumulant generating function (SCGF) $\lambda(k)$.

Given a sequence of codes with increasing code distance $d$, we define the scaled cumulant generating function
\begin{equation}
    \lambda(k)
    \equiv
    \lim_{d\rightarrow\infty}
    \frac{1}{d}
    \log \left\langle e^{k|\Delta F_d|}\right\rangle=\lim_{d\rightarrow\infty} \lambda_d(k),
\end{equation}
where $\Delta F_d$ denotes the free-energy difference associated with the distance-$d$ code. Informally, the G\"artner-Ellis theorem states that if $\lambda(k)$ exists, and is finite and differentiable on an interval containing the origin, then the random variable $|\Delta F_d|/d$ satisfies a large-deviation principle with a convex rate function
\begin{equation*}
    I(s)
    =
    \sup_k \left( ks-\lambda(k)\right),
\end{equation*}
given by the Legendre--Fenchel transform of $\lambda(k)$~\cite{dembo1998large,Touchette2009}. Furthermore, due to the Nishimori-type symmetry $P(\Delta F)=P(-\Delta F)e^{\beta \Delta F}$, the large-deviation properties of $|\Delta F|$ and $\Delta F$ are equivalent. Notably, although our discussion based on the G\"artner--Ellis theorem relies on the Nishimori symmetry, we numerically observe that the LDP also holds for the complementary gap, which serves as a proxy for the free-energy difference $\Delta F$ but does not obey the Nishimori symmetry.

\textbf{Analysis of $\lambda(k)$. }
Prior to the analysis of $\lambda(k)$, we presume several benign but physically reasonable properties of $P(\Delta F_d)$:
\begin{enumerate}
\item Both $P(\Delta F_d)$ and $P(|\Delta F_d|)$ are peaked near the typical value $\langle \Delta F_d\rangle$, and decay monotonically in the left tail $x\le \langle \Delta F_d\rangle$.

\item Below threshold, the logical error rate decays exponentially with code distance,
\[
p_f(d)\simeq e^{-\alpha d}.
\]
\end{enumerate}
We first show that the probability density near $\Delta F_d=0$ exhibits the same exponential scaling as $p_f(d)$. Using the Nishimori relation,
\begin{equation}
    p_f(d)
    =
    \int \dd |\Delta F_d|
    \frac{P(|\Delta F_d|)}{1+e^{|\Delta F_d|}}
    \simeq
    \int \dd |\Delta F_d|
    P(|\Delta F_d|)e^{-|\Delta F_d|},
\end{equation}
hence
\[
\lambda(-1)=-\alpha,
\]
provided the logical error rate indeed obeys exponential scaling.

Since
\[
\frac{P(|\Delta F_d|)}{1+e^{|\Delta F_d|}}
=
P(-|\Delta F_d|),
\]
and $P(\Delta F_d)$ is presumed to decay monotonically in the left tail, the integral above is controlled by the region near $\Delta F_d=0$. Moreover, for any finite $d$, the range of $|\Delta F_d|$ is linearly bounded,
\[
|\Delta F_d|\le Cd,
\]
since the logical error probability of the trivial syndrome is at least $p^d$. Consequently, for any fixed $\epsilon>0$,
\begin{equation}
    \frac12 P(|\Delta F_d|\le \epsilon)e^{-\epsilon}
    \le
    p_f(d)
    \le
    Cd\,\epsilon^{-1}
    P(|\Delta F_d|\le \epsilon)e^{-\epsilon}.
\end{equation}
The lower bound is obtained by restricting the integral to
$|\Delta F_d|\le \epsilon$, while the upper bound follows from
monotonicity of the integrand together with the fact that the
support of $|\Delta F_d|$ has width at most $O(d)$.

Therefore,
\begin{equation}
    P(|\Delta F_d|\le \epsilon)\simeq e^{-\alpha d},
    \qquad \text{for any fixed }\epsilon>0.
\end{equation}

Notice that
\[
P(|\Delta F_d|)
\frac{e^{(1+k)|\Delta F_d|}}{1+e^{|\Delta F_d|}}
\]
remains monotonically decreasing for $k\le -1$. By the same argument,
\begin{equation}
    P(|\Delta F_d|\le\epsilon)e^{k\epsilon}
    \le
    \langle e^{k|\Delta F_d|}\rangle
    \le
    2Cd\,\epsilon^{-1}
    P(|\Delta F_d|\le\epsilon)e^{k\epsilon},
\end{equation}
which implies
\[
\lambda(k)=-\alpha,
\qquad k\le -1.
\]

Moreover, $\lambda_d(k)$ is monotonically increasing in $k$, while $\lambda(0)=0$. Hence
\begin{equation}
    -\alpha-\delta
    \le
    \lambda_d(k)
    \le
    \delta,
    \qquad \text{for any }\delta>0,
\end{equation}
for sufficiently large $d$ and all $-1\le k\le0$.

As the sequence $\lambda_d(k)$ is bounded, both the limit inferior and limit superior exist pointwise for $k\le0$. If we further believe the code family grows in a sufficiently natural manner, then $\lambda_d(k)$ should not oscillate between distinct accumulation points, and pointwise convergence of $\lambda_d(k)$ to $\lambda(k)$ becomes plausible for $k\le0$.

Moreover, if the system exhibits sufficient regularity below threshold --- for example, if $\lambda(k)$ is $C^1$ on $k\le0$ --- then $\lambda(k)$ becomes a convex continuously differentiable function whose derivative on the interval $[-1,0]$ smoothly ranges from
\[
\lambda'(-1)=0
\]
to
\[
\lambda'(0)=\overline{s}
\equiv
\lim_{d\rightarrow\infty}
\frac{\langle \Delta F_d\rangle}{d}.
\]
This is precisely the behavior expected from the desired large deviation principle for $P(|\Delta F_d|)$ within the regime
\[
0\le |\Delta F_d|
\le
\langle \Delta F_d\rangle.
\]